\documentclass[prd,preprint,superscriptaddress,preprintnumbers,eqsecnum,showpacs,nofootinbib,nobibnotes]{revtex4}
\usepackage{amsfonts,amsmath,amssymb,bm,natbib}
\usepackage{graphicx} 
\usepackage{pstricks}

\newcommand{\be}{\begin{equation}}
\newcommand{\bea}{\begin{eqnarray}}
\newcommand{\ee}{\end{equation}}
\newcommand{\eea}{\end{eqnarray}}
\newcommand{\sla}{\slash \hspace{-0.22cm}}

\def\1eq#1{Eq.~(\ref{#1})}

\def\2eqs#1#2{Eqs.~(\ref{#1}) and~(\ref{#2})}
\def\3eqs#1#2#3{Eqs.~(\ref{#1}),~(\ref{#2}) and~(\ref{#3})}
\def\noeq#1{(\ref{#1})}

\def\mrgi{\overline m}
\def\fig#1{Fig.~\ref{#1}}

\def\spr{\!\cdot\!}
\def\eg{{\it e.g.}, }

\def\ie{{\it i.e.}, }

\def\n#1{({\it #1}\,)}

\begin{document}

\title{QCD effective charge 
 from the \\ three-gluon vertex of the background-field method}

\author{D. Binosi}
\affiliation{European Centre for Theoretical Studies in Nuclear
Physics and Related Areas (ECT*) and Fondazione Bruno Kessler, \\Villa Tambosi, Strada delle
Tabarelle 286, 
I-38123 Villazzano (TN)  Italy}

\author{D. Iba\~nez}

\author{J. Papavassiliou}
\affiliation{\mbox{Department of Theoretical Physics and IFIC,  
University of Valencia}
E-46100, Valencia, Spain}

\begin{abstract}

In this  article we study in detail the prospects  of determining the
infrared finite QCD effective charge from a special kinematic limit of
the vertex  function corresponding  to three background  gluons.  This
particular  Green's  function  satisfies  a  QED-like  Ward  identity,
relating it  to the gluon propagator,  with no reference  to the ghost
sector. Consequently,  its longitudinal form factors  may be expressed
entirely  in terms  of the  corresponding gluon  wave  function, whose
inverse  is proportional  to  the effective  charge.  After  reviewing
certain  important  theoretical  properties,  we  consider  a  typical
lattice quantity involving this vertex, and derive its exact dependence
on the various  form factors, for arbitrary momenta.  We then focus on
the particular  momentum configuration that  eliminates any dependence on  the (unknown)  transverse form  factors, projecting  out  only the desired quantity.  A preliminary numerical analysis indicates that the effective   charge  is   relatively  insensitive   to   the  numerical uncertainties  that may  afflict  future simulations of the aforementioned lattice quantity.  The  numerical difficulties associated with a parallel determination of  the dynamical gluon mass are briefly discussed.

\end{abstract}


\maketitle

\section{Introduction}

The ongoing quest for a deeper understanding of the 
nonperturbative features of the basic Green's functions of QCD has been 
benefited considerably by the constructive interaction  
between lattice simulations and the Schwinger-Dyson equations (SDEs).
One of the central issues in this search is the systematic study 
of how the effects of dynamical  
mass generation    
manifest themselves in the  nonperturbative
structure of the propagators of the fundamental 
fields (quarks, gluons, and ghosts).
In particular, 
the concept of a momentum-dependent gluon mass~\cite{Cornwall:1981zr}
has received renewed attention, 
because it provides a natural  explanation 
for the infrared finiteness of the gluon propagator, $\Delta(q^2)$, and ghost 
dressing function, $F(q^2)$, observed in large-volume lattice simulations carried out in the Landau gauge,
both in  $SU(2)$~\cite{Cucchieri:2007md} and in $SU(3)$~\cite{Bogolubsky:2009dc}.

Especially interesting in this context is the definition 
and properties of the QCD effective charge, as well as its interplay with the 
gluon mass.
In fact, as has been explained in detail in a series of works~\cite{Aguilar:2008fh,Aguilar:2009nf,Aguilar:2010gm}, 
if the  finiteness of the aforementioned Green's functions  
is to be attributed  
to the dynamical generation of a gluon mass [{\it i.e.}, $\Delta^{-1}(0)= m^2(0)$], then   
the corresponding effective charge, when properly defined, saturates also  
at a finite, nonvanishing value~\cite{Cornwall:1981zr,Aguilar:2009nf}. 
Within this picture, the so-called ``freezing'' of the charge is a consequence of the 
presence of the gluon mass in the corresponding logarithms, 
which are cured from the perturbative Landau pole, are well-defined for all physical momenta,  
and acquire a finite value at $q^2=0$.  
It is important to emphasize that this characteristic property of the QCD coupling, 
which endows the theory with an infrared fixed point,      
has been advocated by a large number of very distinct  
theoretical and phenomenological approaches, see~\cite{Cornwall:1981zr,Dokshitzer:1995qm,vonSmekal:1997is,Shirkov:1997wi,Aguilar:2002tc,Prosperi:2006hx,Fischer:2006ub,Aguilar:2009nf,Brodsky:2010ur,Courtoy:2013qca} and references therein. 

A self-consistent framework for  the study of the effective charge 
is provided by the fusion of the pinch technique (PT)~\cite{Cornwall:1981zr,Cornwall:1989gv,Pilaftsis:1996fh, 
Binosi:2002ft,Binosi:2003rr,Binosi:2009qm} with  
the background field method (BFM)~\cite{Abbott:1980hw},  
known in the literature as the ``PT-BFM'' scheme~\cite{Aguilar:2006gr,Binosi:2007pi,Binosi:2008qk}.
The natural starting point in this context is the PT gluon propagator, $\widehat\Delta(q^2)$,
which is known to coincide with the two-point function of two background gluons.
Note that $\widehat\Delta(q^2)$ and  $\Delta(q^2)$, as well as their individual  
kinematic components introduced below, are related by powerful identities~\cite{Grassi:1999tp,Binosi:2002ez} [see, \eg \1eq{BQIpropagatorsa}],  
involving an auxiliary two-point function, which, in the deep infrared,  
practically coincides with the ghost dressing function~\cite{Grassi:2004yq,Aguilar:2009pp}. 

In the presence of a dynamically generated mass, the (Euclidean) $\widehat\Delta(q^2)$ 
 assumes the form $\widehat\Delta^{-1}(q^2) =q^2 {\widehat J}(q^2) + {\widehat m}^2(q^2)$,  
where the first term corresponds to the ``kinetic term'', or ``wave function'' contribution, 
while the second denotes the momentum-dependent mass~\cite{Aguilar:2009ke}.
Then,  the (infrared finite) effective charge ${\overline\alpha}(q^2)$ may be defined as  
\mbox{${\overline\alpha}(q^2) = \alpha_s(\mu^2) {\widehat J}^{-1}(q^2,\mu^2)$}, 
where $\mu$ is the renormalization (subtraction) point chosen, within an appropriate renormalization scheme~\cite{Aguilar:2009ke}.
By virtue of the QED-like Ward identities (WIs) satisfied in the PT-BFM framework,
this latter quantity is formally renormalization group (RG)-invariant ($\mu$-independent).   
Note that 
the freezing property of this ${\overline\alpha}(q^2)$ hinges crucially 
on the fact that the infrared finiteness of $\widehat\Delta(q^2)$  
has been accounted for by the inclusion of the  ${\widehat m}^2(q^2)$ in the 
above decomposition; 
instead, the alternative definition in terms of the 
gluon dressing function $\widehat{\cal Z}(q^2) = q^2 \widehat\Delta(q^2)$, 
namely \mbox{${\overline\alpha}(q^2) = \alpha_s(\mu^2) \widehat{\cal Z}(q^2,\mu^2)$},
gives rise to a RG-invariant effective charge that vanishes trivially at the origin~\cite{Aguilar:2009nf}.

Evidently, in order to proceed further with the determination of the physical 
effective charge, one needs 
information on the behavior of  ${\widehat J}(q^2)$. To be sure, ${\widehat J}(q^2)$  
could be obtained 
from \1eq{BQIJ} if its  conventional counterpart, ${J}(q^2)$, were known; however, unlike  $F(q^2)$, this latter 
quantity may not be directly extracted from existing lattice simulations.
Indeed, the lattice determines the full propagator $\Delta(q^2)$, but offers  
no direct information on the momentum dependence of the individual components.
 
Similarly, within the nonperturbative PT-BFM framework, both  ${\widehat J}(q^2)$ and  ${\widehat m}^2(q^2)$ 
satisfy independent (but coupled) non-linear  
integral equations, which are obtained from the SDE of the $\widehat\Delta(q^2)$ 
following a well-defined procedure, developed in a series of recent works~\cite{Aguilar:2011ux,Binosi:2012sj}. 
These two equations may, in principle, determine the complete dynamics of these two quantities.
In practice, however, one is considerably limited  
by the fact 
that the main ingredients entering in them are the (largely unexplored) fully dressed 
three- and four-gluon vertices, for arbitrary values of their momenta, and, as a result, only 
approximate solutions may be obtained. 

It would be clearly desirable, therefore, to determine  ${\widehat J}(q^2)$ 
from an approach that is {\it ab-initio} exact, in the sense that it 
does not involve any field-theoretic approximations.
To that end, in this article we explore the possibility of 
extracting this important quantity from a possible lattice simulation of the 
PT-BFM three-gluon vertex~\cite{Cornwall:1989gv,Binger:2006sj}, 
to be denoted by $\widehat\Gamma$, 
defined as the one-particle irreducible part of the 
correlation function involving three background gluons (the  prospects for a lattice nonperturbative formulation of the BFM have been recently revitalized due to the results of~\cite{Binosi:2012st,Cucchieri:2012ii}). 

The fundamental reason why this vertex can furnish clean information on 
${\widehat J}(q^2)$ is the  
Abelian WI that it satisfies~\cite{Cornwall:1989gv}. Specifically, to all orders in perturbation theory  
this WI involves only the difference of $\widehat\Delta^{-1}(q^2)= q^2 {\widehat J}(q^2)$, defined at the 
appropriate momenta; this is to be contrasted with the usual 
Slavnov-Taylor identity (STI) satisfied by the 
conventional three-gluon vertex~\cite{Marciano:1977su}, which involves, in addition, 
contributions from the ghost-sector of the theory, and especially 
from the so-called ghost-gluon kernel~\cite{Ball:1980ax}.  

Of course, to properly account for mass generation, $\widehat\Gamma$ must be supplemented by 
a special nonperturbative vertex, denoted by $\widehat V$, which 
contains the necessary poles to enforce gauge invariance in the presence of a gluon mass~\cite{Aguilar:2011ux}.
This particular vertex is completely longitudinally coupled, 
and its divergence furnishes precisely the missing mass terms that convert   
$q^2 {\widehat J}(q^2)$ into a massive $\widehat\Delta^{-1}(q^2)$, thus 
maintaining the form of the original WI intact. 
However, 
when the full vertex 
$\widehat\Gamma+\widehat V$ is contracted by three polarization tensors, as happens typically 
in lattice calculations in the Landau gauge~\cite{Cucchieri:2006tf}, any reference on 
the (completely longitudinal)
$\widehat V$ disappears~\cite{Aguilar:2011ux}, 
and only the dependence on the nonperturbative  ${\widehat J}$, 
defined at different momenta scales, survives. 
Then, a special kinematic limit, 
frequently employed in the lattice studies of three-point functions~\cite{Cucchieri:2006tf}, 
converts the lattice quantity of interest
into a function of a single variable, given simply by 
the first derivative of $q^2 {\widehat J}(q^2)$. 
Finally, a straightforward integration of the lattice result over the 
relevant momentum interval furnishes ${\widehat J}(q^2)$, and consequently 
${\overline\alpha}(q^2)$; this constitutes the central result of the present work. 

It is clear that the usefulness of 
the aforementioned exact result  
must be assessed  within the context 
of a realistic lattice simulation, 
taking into account, to some extent,
the practical limitations associated with such an endeavor. 
In particular, 
it is important to provide a rough estimate of the errors that the various numerical 
uncertainties may introduce to the  
effective charge and, subsequently, the gluon mass.
A simple modeling of these effects reveals that 
the predictions  obtained for the 
effective charge are rather robust, and that the induced deviations do 
not alter significantly its theoretically expected behavior. 
Instead, with the exception of the 
deep infrared, the extraction of the 
gluon mass is afflicted by important qualitative discrepancies.

The article  is organized as  follows.  In Sec.~\ref{sec:gen}  we 
define the basic quantities appearing in this problem,
and summarize some important relations, characteristic 
to the PT-BFM framework. 
In Sec.~\ref{sec:RGI} we introduce the 
RG-invariant definition of the QCD
effective charge and of the gluon mass;  in particular, 
we  put forth an
interesting analogy between this latter RG-invariant mass and the standard  
constituent quark  mass. In  Sec.~\ref{sec:3g} we present 
the relevant three-gluon vertex, together  with its basic properties.  
Particular  attention is payed  to  the  linear
(ghost-free)  WI  satisfied  by  this  vertex, which  allows  for  the
complete  determination of  its longitudinal form  factors in
terms  of the gluon propagator,  with  no  reference  to  ghost  Green's
functions. Sec.~\ref{sec:lat} contains the main results of the paper. First, 
we briefly review the  general prospects of simulating PT-BFM Green's  functions on the
lattice.  Then,  assuming that  such  a  simulation  can be actually 
carried out for  the vertex, we show  how the effective charge
and gluon mass  can be directly reconstructed from  the data obtained  
in a commonly used  kinematical limit. 
In addition, we carry
out a  detailed numerical study  on how the unavoidable numerical  errors 
of a possible simulation  
propagate to the  relevant physical quantities.
Our  analysis shows that,
while a faithful approximation of  the effective charge can be generally 
obtained,  the  reconstruction  of  the  gluon running mass  is  much  more
subtle, displaying large  fluctuations due to the inevitable distortion of
delicate numerical cancellations.  Finally, our conclusions  are   presented in 
Sec.~\ref{sec:conc}.

\section{General considerations}\label{sec:gen}

The full gluon propagator 
$i\Delta^{ab}_{\mu\nu}(q)=\delta^{ab}\Delta_{\mu\nu}(q)$ in the Landau gauge is defined as
\be
\Delta_{\mu\nu}(q)=- i P_{\mu\nu}(q)\Delta(q^2),
\label{prop}
\ee 
where
\be
P_{\mu\nu}(q)=g_{\mu\nu}- \frac{q_\mu q_\nu}{q^2} 
\ee
is the usual transverse projector, 
and the scalar cofactor $\Delta(q^2)$  
is related to the (all-order) gluon self-energy $\Pi_{\mu\nu}(q)=P_{\mu\nu}(q)\Pi(q^2)$  through
\be
\Delta^{-1}({q^2})=q^2+i\Pi(q^2).
\label{defPi}
\ee
It is advantageous to introduce the {\it inverse} of the gluon dressing function, $J(q^2)$, defined as~\cite{Ball:1980ax}
\be
\Delta^{-1}({q^2})=q^2 J(q^2).
\label{defJ}
\ee
At tree-level, $J(q^2) =1$.   
Perturbatively, at one-loop, it is given by 
\be
J(q^2) = 1+ \frac{13C_{\rm A}g^2}{96\pi^2} \ln\left(\frac{q^2}{\mu^2}\right),
\label{Jpert}
\ee
where $C_A$ denotes the Casimir eigenvalue of the adjoint representation [$C_A=N$ for $SU(N)$],
and the renormalization has been carried out in the momentum-subtraction (MOM) scheme. 
Evidently, $J^{-1}(q^2)$
displays a Landau pole at $q^2 = \mu^2 \exp(-96\pi^2/13C_{\rm A}g^2)$.

The generation of a dynamical gluon mass leads to the infrared finiteness of the 
gluon propagator, to be denoted by $\Delta_m(q^2)$.
In particular, in Minkowski space one has  
\be
{\Delta}_m^{-1}(q^2)= q^2 {J}_m(q^2)-{m}^2(q^2),
\label{massive}
\ee
with ${m}^2(0) \neq 0$. 
The subscript ``$m$'' in ${J}_m$
indicates that the resulting expressions are regulated by the presence of ${m}^2(q^2)$.
Specifically, after gluon mass generation, the $J_m(q^2)$ may be {\it qualitatively} described by 
\be
J_m(q^2)=  1+ \frac{13C_{\rm A}g^2}{96\pi^2} 
\ln\left(\frac{q^2 +\rho \,m^2(q^2)}{\mu^2}\right),
\label{Jmass}
\ee  
with $g^2$, the constant $\rho$, and $m^2(q^2)$ such that $J_m(q^2)>0$ for all values of $q^2$,   
thus avoiding completely the appearance of a Landau pole.

The detailed dynamics that govern $J_m(q^2)$ and $m^2(q^2)$ are determined by   
two coupled integral equations, obtained from the SDE of the gluon propagator.
Specifically, after a nontrivial reorganization of terms, one obtains  
an inhomogeneous equation for $J_m(q^2)$, and a homogeneous one for $m^{2}(q^2)$,
of the general form~\cite{Aguilar:2011ux,Binosi:2012sj}  
\bea
J_m(q^2) [1 + G(q^2)] &=& 1+ \mu^\epsilon\int\!\frac{\mathrm{d}^dk}{(2\pi)^d}\, {\cal K}_1 (k,q,m^2,\Delta_m),
\nonumber\\
m^{2}(q^2)[1 + G(q^2)] &=&  \mu^\epsilon\int\!\frac{\mathrm{d}^dk}{(2\pi)^d}\,  {\cal K}_2 (k,q,m^2,\Delta_m),
\label{meq}
\eea
where $d=4-\epsilon$ is the space-time dimension and $\mu$ the 't Hooft mass.
Note that  the corresponding kernels are such that, as $q\to 0$,  
${\cal K}_{1,2} (k,q,m^2,\Delta_m)\neq 0$; in fact, in Euclidean space, the solution of these equations
furnishes $J_m(q^2)>0$ and $m^{2}(q^2)>0$, as expected on physical grounds. 

The appearance of the factor $[1 + G(q^2)]$ on the lhs of \1eq{meq} stems from the fact that 
the corresponding equations are not derived from the standard SDE for the gluon propagator, but rather 
from its PT-BFM version; the advantages of this particular approach 
have been explained in detail in a series of articles~\cite{Aguilar:2006gr,Binosi:2007pi,Binosi:2008qk}.  
In the PT-BFM formalism the natural separation of the gluonic field
into a ``quantum'' ($Q$) and a ``background'' ($B$) gives rise to an extended set of
Feynman rules, and leads to an increase in the type of possible Green's functions that 
one may consider. In the case of the gluonic two-point function,  
in addition to the conventional $QQ$ 
gluon propagator, $\Delta$, two additional quantities appear: 
the $QB$ propagator, $\widetilde{\Delta}$, 
mixing one quantum gluon with one background gluon, and the $BB$ propagator, 
$\widehat{\Delta}$, with two background gluon legs. It turns out that 
these three propagators are related by the all-order identities~\cite{Grassi:1999tp,Binosi:2002ez}
\begin{equation}
\Delta(q^2) = [1+G(q^2)]\widetilde{\Delta}(q^2) = [1+G(q^2)]^2\widehat{\Delta}(q^2), 
\label{BQIpropagatorsa}
\end{equation}
usually referred to as Background-Quantum identities (BQIs). 

The function $G(q^2)$ is defined as the $g_{\mu\nu}$ component of the special two-point function 
\bea
\Lambda_{\mu\nu}(q)&=&-i g^2 \mu^{\epsilon} C_A \!\int  \frac{\mathrm{d}^d k}{(2\pi)^{d}} \,
\Delta_\mu^\sigma(k)D(q-k)H_{\nu\sigma}(-q,q-k,k)\nonumber\\
&=&g_{\mu\nu}G(q^2)+\frac{q_\mu q_\nu}{q^2}L(q^2),
\label{Lambda}
\eea
where  
$D^{ab}(q^2)=\delta^{ab}D(q^2)$ is the ghost propagator, and $H_{\nu\sigma}$ is 
the gluon-ghost kernel~\cite{Pascual:1984zb,Davydychev:1996pb}. 
Pertubatively, at one-loop, we have (Landau gauge, MOM scheme) 
\be
1+G(q^2) = 1 +\frac{9}{2}
\frac{C_{\rm {A}}g^2}{96\pi^2}\ln\left(\frac{q^2}{\mu^2}\right).
\label{Gpert}
\ee

Just as the usual quantum gluon propagator, 
the $BB$ propagator   $\widehat{\Delta}$ is also infrared finite, and 
must be parametrized in complete analogy with \1eq{massive}, namely 
\be
\widehat{\Delta}_m^{-1}(q^2)= q^2 \widehat{J}_m(q^2)-\widehat{m}^2(q^2),
\label{hatmassive}
\ee
and an exactly analogous formula  holds for $\widetilde{\Delta}$ (not used here).
Then, one can establish that the BQIs hold individually for the kinetic and mass terms, {\it i.e.},  
\bea
\widehat{J}_m(q^2) &=& [1 + G(q^2)]^2 J_m(q^2),
\label{BQIJ}
\\
{\widehat m}^2(q^2) &=& [1 + G(q^2)]^2 m^2(q^2)
\label{BQImass}
\eea
Use of \1eq{BQIJ}, together with \1eq{Gpert} and \1eq{Jpert}, furnishes the one-loop perturbative 
expression for $\widehat{J}(q^2)$, namely
\be
\widehat{J}(q^2) = 1+ b g^2\ln\left(\frac{q^2}{\mu^2}\right),
\ee
where  $b = 11 C_A/48\pi^2$  is the first coefficient of the QCD $\beta$-function. 
Evidently, as is well-known~\cite{Abbott:1980hw}, the 
propagator $\widehat\Delta(q^2)$ absorbs all the RG logarithms, 
exactly as happens in QED with the photon self-energy.  

\section{\label{sec:RGI} Effective charge and RG-invariant gluon mass}

Let us next recall that, 
due to the Abelian WIs satisfied by the PT-BFM Green's functions,
the renormalization constants of the gauge-coupling and of $\widehat\Delta(q^2)$, 
defined as 
\bea
g(\mu^2) &=&Z_g^{-1}(\mu^2) g_0 ,\nonumber \\
\widehat\Delta(q^2,\mu^2) & = & \widehat{Z}^{-1}_A(\mu^2)\widehat{\Delta}_0(q^2), 
\label{conrendef}
\eea
where the ``0'' subscript indicates bare quantities, satisfy the 
QED-like relation 
\be
{Z}_{g} = {\widehat Z}^{-1/2}_{A}.
\label{ptwi}
\ee
As a result, the product  
\be
{\overline {\!d\,}}_0(q^2) \equiv g^2_0 \widehat\Delta_0(q^2) = 
g^2 \widehat\Delta(q^2) \equiv {\overline {\!d\,}} (q^2),
\ee
forms a RG invariant  \mbox{($\mu$-independent)} quantity.
As has been explained in the recent literature~\cite{Aguilar:2009ke}, ${\overline {\!d\,}} (q^2)$ 
may be cast in the form 
\be
{\overline {\!d\,}}(q^2) = \frac{{\overline g}^2(q^2)}{q^2 + \mrgi^2 (q^2) },
\ee
with
\be
{\overline g}^2 (q^2) = g^2 {\widehat J}_m^{-1} (q^2),
\label{effch}
\ee
and 
\be
{\mrgi}^2 (q^2) = {\widehat m^2} (q^2) {\widehat J}_m^{-1} (q^2).
\label{RGmass}
\ee
Note that the two quantities defined above are individually RG invariant .
The usual effective charge, ${\overline\alpha}(q^2)$, is obtained from \1eq{effch}
simply as ${\overline\alpha}(q^2)\equiv {\overline g}^2 (q^2)/4 \pi$. 
At one-loop,
\be
\overline{g}^2(q^2) = \frac{g^2}{1+  b g^2\ln\left(q^2/\mu^2\right)}
= \frac{1}{b\ln\left(q^2/\Lambda^2\right)}.
\label{effch1loop}
\ee
where $\Lambda$ denotes an RG invariant  mass scale of a few hundred ${\rm MeV}$.

It is interesting to observe the analogy between the RG invariant  mass defined in 
\1eq{RGmass} and the corresponding constituent quark mass, familiar from a plethora  
of studies on chiral symmetry breaking. Specifically,  
the quark propagator is usually cast in the form 
\bea
S^{-1}(p) &=&  A(p^2)\,\sla{p} - B(p^2) \mathbb{I} 
\nonumber\\
&=& A(p^2)[\sla{p}-{\mathcal{M}}(p^2) \mathbb{I}],
\label{qpropAB}
\eea
where $\mathbb{I}$ is the identity matrix, and the 
term \mbox{$A^{-1}(p^2)$} is often referred to in the literature  as the ``quark wave function''.
Note that, the quark acquires a dynamical mass (signaling the breaking of 
chiral symmetry) provided that $B(p^2)$ is different from zero, and that the constituent quark mass, 
${\mathcal{M}}(p^2)=B(p^2)/A(p^2)$, is RG invariant . Then, it is natural to propose an analogy between 
the gluon and quark propagators; 
clearly, ${\widehat J}_m(q^2)$ corresponds to $A(p^2)$, 
while  ${\widehat m}^2$ plays exactly the role of $B(p^2)$. 
Then, the division by ${\widehat J}_m(q^2)$ and $A(p^2)$ gives rise, in both cases, to RG invariant  masses, 
suggesting a close correspondence between ${\mrgi}^2(q^2)$ and  ${\mathcal{M}}(p^2)$.

Next, using the BQIs~(\ref{BQIJ}) and (\ref{BQImass}) to relate the components of $\widehat\Delta_m(q^2)$ 
to the corresponding ones of $\Delta_m(q^2)$, we get 
\be
{\widehat m}^2 (q^2) {\widehat J}_m^{-1} (q^2) = {m}^2 (q^2) {J}_m^{-1} (q^2), 
\ee
which finally furnishes a set of relations equivalent to \noeq{effch} and \noeq{RGmass}, 
\bea
{\overline g}^2 (q^2) &=& g^2 [1+G(q^2)]^{-2} {J}_m^{-1} (q^2),
\label{effchG}\\
{\mrgi}^2 (q^2) &=& {\widehat m^2} (q^2)[1+G(q^2)]^{-2} {J}_m^{-1} (q^2).
\label{RGmassG}
\eea

The basic \2eqs{effch}{RGmass}, or alternatively \2eqs{effchG}{RGmassG}, 
express the fundamental quantities  ${\overline g}^2 (q^2)$ and ${\mrgi}^2 (q^2)$
in terms of ${\widehat J}_m(q^2)$, or $J_m(q^2)$ and $G(q^2)$, respectively.
It is therefore important to review the state-of-the-art in the determination of these 
quantities, both in the continuum as well as on the lattice.

To that end, let us first focus on the quantities $J_m(q^2)$ and $G(q^2)$, defined 
in the context of the conventional covariant gauges; in fact, we will specialize the 
discussion in the Landau gauge, where lattice simulations are usually performed.
 
Let us say from the outset that lattice simulations of the gluon propagator ${\Delta}_m(q^2)$ 
cannot furnish $J_m(q^2)$ without any additional input, for the simple
reason that they provide the entire combination of $J_m(q^2)$ and ${m^2} (q^2)$, as appears 
in \1eq{massive}, but not the individual components comprising it. 
The SDEs, on the other hand, when appropriately reorganized, give rise to 
two coupled integral equations, one for $J_m(q^2)$ and one for ${m^2} (q^2)$, 
as shown schematically in \1eq{meq}. In principle these two equations should 
furnish the exact behavior of $J_m(q^2)$ and ${m^2} (q^2)$; in practice, one is limited 
by the fact that the closed form of the 
kernels ${\cal K}_1$ and  ${\cal K}_2$ are only approximately known. In fact, 
${\cal K}_2$ is better known than ${\cal K}_1$, mainly because 
the only unknown quantity that enters in  ${\cal K}_2$ 
is the full three-gluon vertex $\Gamma_{\alpha\mu\nu}$, whereas 
in  ${\cal K}_1$ appears, in addition, the full four-gluon vertex. Even though neither of these two vertices is 
known, the STIs that the three-gluon vertex satisfies [see \1eq{WIQ}] furnish 
nontrivial information on the structure of its longitudinal components, thus providing certain 
field-theoretically motivated approximations for ${\cal K}_2$; 
on the contrary, no such construction exists for the four-gluon vertex, which  
remains practically unexplored. 

The situation described above has motivated the advent of an approach which combines 
the information from both the lattice and the SDEs. In particular, 
instead of solving the system of \1eq{meq}, only the mass equation 
(involving the better known ${\cal K}_2$) is solved, using as input in it 
the lattice data for ${\Delta}_m(q^2)$. Then, once a  
solution for ${m^2} (q^2)$ has been obtained, one can use \1eq{massive} 
to extract the approximate form of $J_m(q^2)$.    
   
Regarding the function $G(q^2)$, notice  that, quite remarkably, in the Landau gauge it coincides with the so-called Kugo-Ojima function~\cite{Grassi:2004yq,Aguilar:2009pp}. In addition, there is a deep connection between the form factors $G(q^2)$ and $L(q^2)$ and the ghost dressing function
\be
F(q^2)=q^2D(q^2),
\ee
expressed through the identity~\cite{Grassi:2004yq,Aguilar:2009pp}
\be \label{funrel}
1+G(q^2)+L(q^2)=F^{-1}(q^2). 
\ee
Since in $d=4$ it is known that $L(q^2)\ll G(q^2)$ over the entire momentum range~\cite{Aguilar:2009pp}, one finally arrives at the result 
\be
1+G(q^2)\approx F^{-1}(q^2),
\ee 
which becomes an exact relation at $q^2=0$, since in this case $L(0)=0$.

From the above discussion it is clear that one cannot obtain direct information on $J_m(q^2)$ 
from the gluonic two point sector. Turning to the three-point functions, 
one might think that, given that $J_m(q^2)$  enters into the longitudinal components of 
the conventional three-gluon vertex, as first demonstrated by Ball and Chiu~\cite{Ball:1980ax}, a judicious 
combination of them could project it out. However, the problem in this case is the 
``contamination'' from the various form-factors comprising $H_{\mu\nu}$, which enter nontrivially in the 
various expressions. 

It turns out that the PT-BFM three-gluon vertex lends itself for this type of analysis.
Given that it satisfies QED-like WIs instead of STIs, there is no ghost sector contributions:
the longitudinal form factors can be expressed in terms of the ${\widehat J}_m$ only.    
To be sure, the transverse form factors enter in general, and are unknown.
But, as we will see in Sec.~\ref{sec:lat}, there is a simple kinematic limit, 
studied usually in the lattice simulations of the (conventional) vertex, which 
eliminates completely the transverse components, and projects out solely 
the first derivative of $q^2 {\widehat J}_m(q^2)$.
It is therefore clear that a possible lattice extraction  of ${\widehat J}_m(q^2)$ would provide  
an immediate determination of ${\overline g}^2 (q^2)$ from \1eq{effch}.

On the other hand, 
regarding the prospects of extracting the RG invariant  gluon mass from \1eq{RGmass} or \noeq{RGmassG},
one may envisage two basic scenarios 
[again assuming independent knowledge of ${\widehat J}_m(q^2)]$:
 
\begin{itemize}

\item[\n{i}]  $\widehat\Delta_m(q^2)$  has been simulated on the lattice. Then, it is direct
to extract ${\widehat m^2} (q^2)$ from \1eq{hatmassive}, since (Euclidean space) 
\be
{\widehat m^2} (q^2) = \widehat\Delta_m^{-1}(q^2) - q^2 {\widehat J}_m(q^2),
\label{latm1}
\ee
or, using \1eq{RGmass},
\be
{\mrgi}^2 (q^2) = \widehat\Delta_m^{-1}(q^2) {\widehat J}_m^{-1}(q^2) - q^2.
\label{latm2}
\ee

\item[\n{ii}]  $\widehat\Delta_m(q^2)$ 
has not been simulated on the lattice. Then, to proceed further, 
one must necessarily employ the BQIs; this is rather feasible,
because, as mentioned earlier, the function $G(q^2)$ has been simulated on the lattice. 
Specifically, one may use the conventional $\Delta_m(q^2)$ obtained from the lattice  
together with $G(q^2)$ to build  $\widehat\Delta_m(q^2)$ by means of the BQI.
Then one may return to the previous case \n{i}, and substitute it into \2eqs{latm1}{latm2}.

\end{itemize} 

This general discussion motivates a systematic study of the  PT-BFM three-gluon vertex. To that end,  in the 
next section we will present a brief reminder on the structure and general properties of the $B^3$ vertex, while  in Sec.~\ref{sec:lat} after summarizing the current prospects of simulating nonperturbative PT-BFM Green's functions on the lattice, we will consider how ${\widehat J}(q^2)$ maybe extracted from a possible lattice simulation of this vertex. 

\section{The PT-BFM three-gluon vertex}\label{sec:3g}

The gauge-invariant three-gluon vertex $\widehat\Gamma_{\alpha\mu\nu}$ 
has been first considered in~\cite{Cornwall:1989gv}, 
where its one-loop construction was carried out by means of the PT, 
and its basic WI was derived (see also~\cite{Binosi:2009qm}). 
It was further studied in~\cite{Binger:2006sj}, with particular emphasis on the 
special relations between gluonic, fermionic, and scalar loop contributions,    
and has been revisited very recently in~\cite{Ahmadiniaz:2012xp},  using string-inspired techniques.
\subsection{General properties}

The equivalence between PT (or the ``generalized PT'') and BFM allows for a concise 
field-theoretic definition of this vertex, as vacuum expectation value 
of the time-ordered product of  three background gluons.
In particular, denoting the full connected three-point function by (momentum space) 
\be
{\widehat{\cal G}}_{\alpha\mu\nu}^{abc}(q,r,p) = 
\frac{\left\langle 0 \left\vert T [\widehat{A}^a_{\alpha}(q) \widehat{A}^b_{\mu}(r) \widehat{A}^c_{\nu}(p)]\right\vert 0\right\rangle}{\langle 0\vert 0\rangle},
\label{3pdef}
\ee
one defines
\be
{\widehat{\cal G}}_{\alpha\mu\nu}^{abc}(q,r,p) = 
\widehat\Delta^{aa'}_{\alpha\alpha'}(q) 
\widehat\Delta^{bb'}_{\mu\mu'}(r) 
\widehat\Delta^{cc'}_{\nu\nu'} (p)  
\widehat\Gamma_{\alpha'\mu'\nu'}^{a'b'c'}(q,r,p), 
\label{ampdef}
\ee
where $\widehat\Gamma_{\alpha\mu\nu}(q,r,p)$ is the amputated three-point function. 

\begin{figure}[!t]
\mbox{}\hspace{-1cm}\includegraphics[scale=0.68]{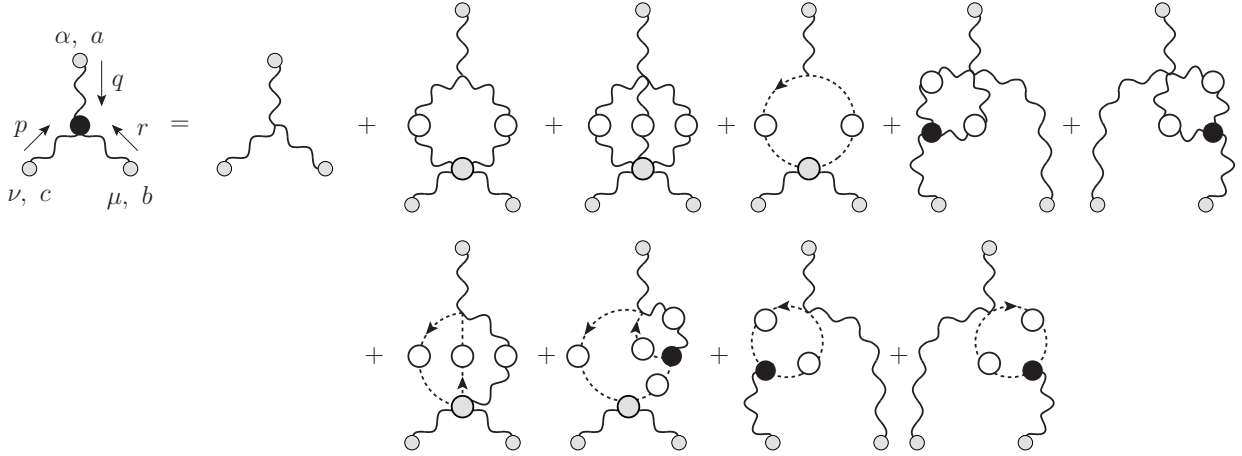}  
\caption{\label{BBB-vertex}The one-particle irreducible $B^3$ vertex and the SDE it satisfies. Background legs are indicated by the small circle at their end; black circles correspond to 1PI Green's functions, white circles represent connected functions, while gray circles indicate SDE kernels. The corresponding symmetry factors may be found in~\cite{Binosi:2008qk}.}
\end{figure}

Exactly as happens with the conventional three-gluon vertex $\Gamma_{\alpha\mu\nu}$, the 
$\widehat\Gamma_{\alpha\mu\nu}$ is fully Bose symmetric. This is to be contrasted with 
the $BQ^2$ vertex, usually denoted by $\widetilde\Gamma_{\alpha\mu\nu}$, which is 
Bose symmetric only under the interchange of its two quantum legs.  
The SDE satisfied by  $\widehat\Gamma_{\alpha\mu\nu}$ is shown in~\fig{BBB-vertex}.
Note that the tree-level expressions for $\Gamma_{\alpha\mu\nu}$ and $\widehat\Gamma_{\alpha\mu\nu}$ coincide:
\be
\Gamma_{\alpha\mu\nu}^{(0)}(q,r,p) = \widehat\Gamma_{\alpha\mu\nu}^{(0)}(q,r,p) = 
(q-r)_{\nu}g_{\alpha\mu} + (r-p)_{\alpha}g_{\mu\nu} + (p-q)_{\mu} g_{\alpha\nu}.
\ee


For the purposes of the present work, 
the most important property of $\widehat\Gamma_{\alpha\mu\nu}$ 
is the fact that it satisfies the completely Bose-symmetric set of Abelian WIs
\bea
q^\alpha\widehat\Gamma_{\alpha\mu\nu}(q,r,p) = p^2 {\widehat J} (p^2)P_{\mu\nu}(p)-r^2 {\widehat J}(r^2)P_{\mu\nu}(r), \nonumber \\
r^\mu\widehat\Gamma_{\alpha\mu\nu}(q,r,p) = q^2 {\widehat J} (q^2)P_{\alpha\nu}(q)-p^2 {\widehat J}(p^2)P_{\alpha\nu}(p), \nonumber \\
p^\nu\widehat\Gamma_{\alpha\mu\nu}(q,r,p) = r^2 {\widehat J} (r^2)P_{\alpha\mu}(r)-q^2 {\widehat J}(q^2)P_{\alpha\mu}(q).
\label{BBBWI}
\eea
These simple WIs are to be contrasted with the STIs satisfied by 
$\Gamma_{\alpha \mu \nu}$, namely
\be 
q^\alpha\Gamma_{\alpha\mu\nu}(q,r,p)=F(q^2)\left[p^2 J(p^2)P_\nu^\alpha(p)H_{\alpha\mu}(p,q,r)-
r^2 J(r^2)P_\mu^\alpha(r) H_{\alpha\nu}(r,q,p)\right]
\label{WIQ}
\ee
and cyclic permutations. The tensorial decomposition of $H_{\nu\mu}$ is given by~\cite{Ball:1980ax} 
\be
H_{\nu\mu}(p,r,q) = g_{\mu\nu}a_{qrp}-r_\mu q_\nu b_{qrp}+q_\mu p_\nu c_{qrp}+q_\nu p_\mu d_{qrp}+p_\mu
p_\nu e_{qrp}, 
\label{Hdec}
\ee
where $a_{qrp}$ is short-hand notation for $a(q,r,p)$, etc.

An immediate consequence of \1eq{BBBWI}
is the QED-like relation  
\be
{\widehat Z}_1 = {\widehat Z}_{A}.
\label{ptwii}
\ee
between the wave-function renormalization for $\widehat\Delta$, introduced in \1eq{conrendef}, and the 
vertex renormalization defined as 
\be
{\widehat Z}_1 \widehat\Gamma^{\alpha\mu\nu}(q,r,p) = \widehat\Gamma^{\alpha\mu\nu}_{R}(q,r,p) .
\ee

In addition, one may extract from the set of WIs given in \1eq{BBBWI} the expression of the vertex 
$\widehat{\Gamma}_{\alpha\mu\nu}$ in the kinematical limit $r\rightarrow 0$, which will be the relevant 
momentum configuration in the subsequent analysis (for a related analysis, see also~\cite{Cornwall:2012mk}). To show that, consider 
the Taylor expansion of a function $f(q,r,p)$ around $r=0$ (and $p=-q$). In general we have
\be
f(q,r,p) = f(q,0,-q) + r^\mu \bigg\lbrace\frac{\partial}{\partial r^\mu}f(q,r,p)\bigg\rbrace_{r=0} + {\cal O}(r^2),
\label{gneralTayl}
\ee
where the Lorentz structure of the function $f$ has been suppressed. Specializing this result to the second WI in \1eq{BBBWI}, one finds
\bea
r^\mu\widehat\Gamma_{\alpha\mu\nu}(q,r,p) &=& r^\mu\widehat\Gamma_{\alpha\mu\nu}(q,0,-q) + {\cal O}(r^2)\nonumber \\
&=&-r^\mu\bigg\lbrace\frac{\partial}{\partial r^\mu}\left[(q+r)^2 {\widehat J} (q+r)P_{\alpha\nu}(q+r)\right]\bigg\rbrace_{r=0} + {\cal O}(r^2).
\label{Taylrhs}
\eea
where we have use the fact that the zero-th order term of~\1eq{gneralTayl}  vanishes in this case.
Thus, equating the coefficients of the terms linear in $r^\mu$, one obtains the relation
\bea
\widehat\Gamma_{\alpha\mu\nu}(q,0,-q) &=& -\bigg\lbrace\frac{\partial}{\partial r^\mu}\left[(q+r)^2 {\widehat J} (q+r)P_{\alpha\nu}(q+r)\right]\bigg\rbrace_{r=0}\nonumber \\
&=& -\frac{\partial}{\partial q^\mu}\left[q^2{\widehat J}(q^2)P_{\alpha\nu}(q)\right].
\label{vertexder}
\eea
Interestingly enough, the above relation is formally equivalent to the well-known QED text-book result
\be
\Gamma_\mu(0,-p,p) = - \frac{\partial}{\partial p^\mu}S^{-1}(p),
\label{QEDvertex}
\ee
obtained from the WI
\be
q^\mu \Gamma_\mu(q,-q-p,p) = S^{-1}(p) - S^{-1}(q+p),
\ee
relating the photon-electron vertex with the electron propagator.

Next, the derivative of \1eq{vertexder} can be easily evaluated by applying the formula
\bea
\frac{\partial}{\partial q^\mu} P_{\alpha\nu}(q) &=& 2q_\mu\frac{q_\alpha q_\nu}{q^4} - \frac{1}{q^2}(g_{\mu\alpha}q_\nu + g_{\mu\nu}q_\alpha) \nonumber \\
&=& -\frac{1}{q^2}[q_\alpha P_{\mu\nu}(q) + q_\nu P_{\mu\alpha}(q)],
\label{Pder}
\eea
yielding the final result
\bea
\widehat\Gamma_{\alpha\mu\nu}(q,0,-q) &=& {\widehat J} (q^2)(g_{\alpha\mu}q_\nu + g_{\nu\mu}q_\alpha - 2g_{\alpha\nu}q_\mu) - 2q^2{\widehat J}'(q^2)q_\mu P_{\alpha\nu}(q) \nonumber \\
&=& {\widehat J} (q^2)\widehat\Gamma_{\alpha\mu\nu}^{(0)}(q,0,-q) - 2q^2{\widehat J}'(q^2)q_\mu P_{\alpha\nu}(q),
\label{vertexrzero}
\eea
where the prime indicates the derivative with respect to $q^2$. Note that, due to Bose symmetry, the procedure described above can be applied exactly in the same way 
for the remaining WIs, in order to obtain the kinematical limits $q,p\rightarrow 0$ of the vertex 
$\widehat{\Gamma}_{\alpha\mu\nu}$. Also, we observe that the tensorial structure of the full vertex $\widehat\Gamma_{\alpha\mu\nu}$ in this particular kinematical limit is not exhausted by the term proportional to the tree level expression. 

\subsection{The pole part of the three-gluon vertex}

As has been explained in a series of recent works, a crucial condition for  obtaining 
an  infrared finite gluon propagator,  without interfering with the 
gauge invariance (or the BRST symmetry) 
of the theory, is the existence of a set of special vertices, to be generically denoted by $V$, 
that are completely longitudinal and contain massless poles.
The dynamical origin 
of the aforementioned poles is purely nonperturbative: for sufficiently strong binding, 
the mass of certain (colored) bound states 
may be reduced to zero~\cite{Jackiw:1973tr,Jackiw:1973ha,Cornwall:1973ts,Eichten:1974et,Poggio:1974qs}. 

The role of these vertices is two-fold. On the 
one hand, thanks to the massless poles they contain, they make possible the emergence of an 
infrared finite solution out of the SDE governing the gluon propagator; thus, one invokes   
essentially a non-Abelian realization of the well-known 
Schwinger mechanism~\cite{Schwinger:1962tn,Schwinger:1962tp}.  
On the other hand, these same poles act like composite Nambu-Goldstone  excitations,  
preserving the form of the STIs of the theory in the presence of a gluon mass.

Specifically, in order for the WIs to maintain the same form before and after 
mass generation, the effective substitution
\be
{\widehat \Delta}^{-1}(q^2) = q^2 {\widehat J}(q^2) \longmapsto \widehat{\Delta}_m^{-1}(q^2)= q^2 \widehat{J}_m(q^2)-\widehat{m}^2(q^2),
\label{massiveprop}
\ee
implemented by the mass generation at the level the gluon propagator, 
must be accompanied by the 
simultaneous replacement of the vertex~\cite{Aguilar:2011ux}  
\begin{equation}
\label{modifiedvertex}
\widehat\Gamma \longmapsto \widehat\Gamma' = \widehat\Gamma\!_m + {\widehat V}.
\end{equation}
Then, since  
\bea
q^\alpha\widehat\Gamma_{\!m\, \alpha\mu\nu}(q,r,p) &=& p^2 {\widehat J}_m (p^2)P_{\mu\nu}(p)-r^2 {\widehat J}_m(r^2)P_{\mu\nu}(r),
\label{WIGm}\\
q^\alpha\widehat{V}_{\alpha\mu\nu}(q,r,p)&=& {\widehat m}^2(r^2)P_{\mu\nu}(r)-{\widehat m}^2(p^2)P_{\mu\nu}(p),
\label{WIV}
\eea
one finds that the corresponding WI satisfied by $\widehat\Gamma'$ would read  
\bea
q^{\alpha}\widehat\Gamma'_{\alpha\mu\nu}(q,r,p) &=& 
q^{\alpha}\left[\widehat{\Gamma}_{m}(q,r,p) + \widehat{V}(q,r,p)\right]_{\alpha\mu\nu}
\nonumber\\
&=& [p^2 {\widehat J}_m (p^2) - {\widehat m}^2(p^2)]P_{\mu\nu}(p) - 
[r^2 {\widehat J}_m (r^2) - {\widehat m}^2(r^2)]P_{\mu\nu}(r)
\nonumber\\
&=& \widehat\Delta_m^{-1}({p^2})P_{\mu\nu}(p) - \widehat\Delta_m^{-1}({r^2})P_{\mu\nu}(r) ,
\label{winpfull}
\eea
which is indeed the first of the identities in \1eq{BBBWI}, with the aforementioned replacement 
\mbox{$\widehat\Delta^{-1} \to \widehat\Delta_m^{-1}$} enforced.

The closed expression for $\widehat{V}_{\alpha\mu\nu}$ may be reconstructed from the WIs it satisfies 
(namely \1eq{WIV} and its cyclic permutations), together with the condition of 
complete longitudinality, \ie 
\be
P^{\alpha'\alpha}(q) P^{\mu'\mu}(r) P^{\nu'\nu}(p) {\widehat V}_{\alpha'\mu'\nu'}(q,r,p)  = 0;
\label{totlon}
\ee 
it reads, 
\be
\widehat{V}_{\alpha\mu\nu}(q,r,p) = \frac{q_\alpha}{q^2}
[{\widehat m}^2(r^2)-{\widehat m}^2(p^2)]P_\mu^\rho(r)P_{\rho\nu}(p) + {\rm c.p.} , 
\ee 
where ${\rm c.p.}$ denotes ``cyclic permutations''. 
As we will see in the next section, due to its completely 
longidudinal nature [{\it viz.} \1eq{totlon}], the contribution  
of this pole vertex to the typical lattice ratio, given in  \1eq{Rprojection}, 
vanishes in the Landau gauge.

\subsection{The longitudinal form factors}

The complete closed form of  $\widehat{\Gamma}_{m}$ is not known; its longitudinal part, however,
may be reconstructed from the WIs that  $\widehat{\Gamma}_{m}$ satisfies, following rather standard 
procedures~\cite{Ball:1980ax}.
Specifically, one begins by separating the vertex into the 
``longitudinal'' and the (totally) ``transverse'' parts,   
\be
\widehat\Gamma_m^{\alpha\mu\nu}(q,r,p)= 
\widehat\Gamma_{m (\ell)}^{\alpha\mu\nu}(q,r,p) + \widehat\Gamma_{m (t)}^{\alpha\mu\nu}(q,r,p), 
\label{decomp}
\ee
where the component $\widehat\Gamma_{m (\ell)}$ satisfies the 
WI of \1eq{WIGm} (and its permutations), whereas  
$q_{\alpha}\widehat\Gamma_{m (t)}^{\alpha\mu\nu}(q,r,p) = 
r_{\mu}\widehat\Gamma_{m (t)}^{\alpha\mu\nu}(q,r,p)=
p_{\nu} \widehat\Gamma_{m (t)}^{\alpha\mu\nu}(q,r,p) = 0$. 

The longitudinal part is then decomposed into 10 form factors $\widehat{X}_i$, according to
\be
\widehat\Gamma_{m (\ell)}^{\alpha\mu\nu}(q,r,p)=\sum_{i=1}^{10}\widehat{X}_i(q,r,p)\ell_i^{\alpha\mu\nu},
\label{tenlon}
\ee
with the explicit form of the tensors $\ell^i$ given by~\cite{Binger:2006sj}
\be
\begin{tabular}{lll}
$\ell_1^{\alpha\mu\nu} =  (q-r)^{\nu} g^{\alpha\mu}$
& 
$\ell_2^{\alpha\mu\nu} =  - p^{\nu} g^{\alpha\mu}$\hspace{.75cm}
&
$\ell_3^{\alpha\mu\nu} =  (q-r)^{\nu}[q^{\mu} r^{\alpha} -  (q\cdot r) g^{\alpha\mu}] $\\
$\ell_4^{\alpha\mu\nu} = (r-p)^{\alpha} g^{\mu\nu}$
&
$\ell_5^{\alpha\mu\nu} =  - q^{\alpha} g^{\mu\nu}$
&
$\ell_6^{\alpha\mu\nu} =  (r-p)^{\alpha}[r^{\nu} p^{\mu} -  (r\cdot p) g^{\mu\nu}]$
\\
$\ell_7^{\alpha\mu\nu} =  (p-q)^{\mu} g^{\alpha\nu}$
&
$\ell_8^{\alpha\mu\nu} = - r^{\mu} g^{\alpha\nu}$
&
$\ell_9^{\alpha\mu\nu} = (p-q)^{\mu}[p^{\alpha} q^{\nu} -  (p\cdot q) g^{\alpha\nu}]$
\\
$\ell_{10}^{\alpha\mu\nu} = q^{\nu}r^{\alpha}p^{\mu} + q^{\mu}r^{\nu}p^{\alpha}$. & &
\end{tabular}
\label{Ls}
\ee
Then, the WI of \1eq{WIGm} and its permutations give rise to an algebraic system 
for the $\widehat{X}_i$, whose solution reads, 
\bea \label{factorsBBB}
&& \widehat{X}_1 = \frac{1}{2}[\widehat{J}_m(q^2) + \widehat{J}_m(r^2)],\quad \widehat{X}_2 = \frac{1}{2}[\widehat{J}_m(q^2) - \widehat{J}_m(r^2)],\quad \widehat{X}_3 = \frac{\widehat{J}_m(q^2) - \widehat{J}_m(r^2)}{q^2 - r^2}, \nonumber \\
&& \widehat{X}_4 = \frac{1}{2}[\widehat{J}_m(r^2) + \widehat{J}_m(p^2)],\quad \widehat{X}_5 = \frac{1}{2}[\widehat{J}_m(r^2) - \widehat{J}_m(p^2)],\quad \widehat{X}_6 = \frac{\widehat{J}_m(r^2) - \widehat{J}_m(p^2)}{r^2 - p^2}, \nonumber \\
&& \widehat{X}_7 = \frac{1}{2}[\widehat{J}_m(p^2) + \widehat{J}_m(q^2)],\quad \widehat{X}_8 = \frac{1}{2}[\widehat{J}_m(p^2) - \widehat{J}_m(q^2)],\quad \widehat{X}_9 = \frac{\widehat{J}_m(p^2) - \widehat{J}_m(q^2)}{p^2 - q^2}, 
\nonumber \\
&& \widehat{X}_{10} = 0.
\eea
Thus, the longitudinal form factors of $\widehat\Gamma_{m (\ell)}^{\alpha\mu\nu}(q,r,p)$ involve {\it only} the quantity $\widehat{J}_m$, since it is the only quantity that enters on the rhs of \1eq{WIGm}.
Instead,  
the corresponding expressions for the form factors of the conventional ($Q^3$) vertex, $\Gamma^{\alpha\mu\nu}$, first 
derived in~\cite{Ball:1980ax}, contain, in addition, the ghost dressing function $F$ and the various form-factors comprising the gluon-ghost kernel $H_{\mu\nu}$.
For example, the closed form of ${X}_7$ (to be employed in the next section) is given by  
\begin{eqnarray}\label{X7}
X_7(q,r,p) &=& \frac{1}{4}\big\lbrace 2[F(q)J_m(p)a_{rqp} + F(p)J_m(q)a_{rpq}] + r^2[F(p)J_m(r)b_{qpr} + F(q)J_m(r)b_{pqr}] \nonumber \\
&+& (q^2 - p^2)[F(r)J_m(q)b_{prq} + F(q)J_m(p)b_{rqp} - F(r)J_m(p)b_{qrp} - F(p)J_m(q)b_{rpq}] \nonumber \\
&+& 2(qr)F(p)J_m(q)d_{rpq} + 2(rp)F(q)J(p)d_{rqp} \big\rbrace.
\end{eqnarray}
Note in addition, that, unlike $\widehat{X}_{10}$, the corresponding ${X}_{10}$ does not vanish. 

Finally, the (undetermined) transverse part of the vertex is described 
by the remaining 4 form factors $\widehat{Y}_i$, 
\be
\widehat\Gamma_{m (t)}^{\alpha\mu\nu}(q,r,p) = \sum_{i=1}^{4}\widehat{Y}_i(q,r,p)t_i^{\alpha\mu\nu},
\label{vtr}
\ee
with the completely transverse tensors $t^i$ given by
\bea
t_1^{\alpha\mu\nu} &=&  
[(q\cdot r) g^{\alpha\mu} - q^{\mu}  r^{\alpha}]
[(r\cdot p) q^{\nu} - (q\cdot p) r^{\nu}]
\nonumber\\
t_2^{\alpha\mu\nu} &=&  
[(r\cdot p) g^{\mu\nu} - r^{\nu}p^{\mu}]
[(p\cdot q) r^{\alpha} - (r\cdot q) p^{\alpha}]
\nonumber \\
t_3^{\alpha\mu\nu} &=&  
[(p\cdot q) g^{\nu\alpha} - p^{\alpha}q^{\nu}]
[(q\cdot r) p^{\mu} - (r\cdot p) q^{\mu}]
\nonumber\\
t_4^{\alpha\mu\nu}&=&g^{\mu\nu}[ (p\cdot q)r^\alpha-(r\cdot q)p^\alpha ]+g^{\alpha\mu}[(r\cdot p)q^\nu-(q\cdot p)r^\nu ] +g^{\alpha\nu}[(r\cdot q)p^\mu -(r\cdot p)q^\mu]\nonumber \\
&+&p^\alpha q^\mu r^\nu-r^\alpha p^\mu q^\nu.
\label{Ts}
\eea

It turns out that in the limit $r\rightarrow 0$ all the transverse tensors in 
\1eq{Ts} are zero, so the transverse part of the vertex vanishes in this limit. On the other hand, 
for the same limit, only the following longitudinal tensors given in \1eq{Ls} survive
\be
\begin{tabular}{lll}
$\ell_1^{\alpha\mu\nu} =  q^{\nu} g^{\alpha\mu}$
& 
$\ell_2^{\alpha\mu\nu} =  q^{\nu} g^{\alpha\mu}$\hspace{.75cm}
&
$\ell_4^{\alpha\mu\nu} =  q^{\alpha}g^{\mu\nu} $\\
$\ell_5^{\alpha\mu\nu} = -q^{\alpha} g^{\mu\nu}$\hspace{.75cm}
&
$\ell_7^{\alpha\mu\nu} =  -2q^{\mu} g^{\alpha\nu}$\hspace{.75cm}
&
$\ell_9^{\alpha\mu\nu} =  -2q^2q^{\mu}P^{\alpha\nu}(q)$,
\end{tabular}
\label{Lsrzero}
\ee
with the associated form factors
\bea \label{factorsBBBzero}
&& \widehat{X}_1 = \frac{1}{2}[\widehat{J}(q^2) + \widehat{J}(0)],\quad \widehat{X}_2 = \frac{1}{2}[\widehat{J}(q^2) - \widehat{J}(0)],\quad \widehat{X}_4 = \frac{1}{2}[\widehat{J}(0) + \widehat{J}(q^2)], \nonumber \\
&& \widehat{X}_5 = \frac{1}{2}[\widehat{J}(0) - \widehat{J}(q^2)],\quad \widehat{X}_7 = \widehat{J}(q^2),\quad \widehat{X}_9 = \widehat{J}'(q^2),
\eea
corresponding to the limit $r\rightarrow 0$ of those appearing in \1eq{factorsBBB}. Thus, using 
these results, it is elementary to show that one is able to reproduce the expression given in 
\1eq{vertexrzero}.

\section{\label{sec:lat}Lattice prospects}

In this section we first discuss the theoretical possibilities 
of simulating PT-BFM Green's  functions on the
lattice.  Then, we consider the relevant lattice quantity, 
and derive its general expression in terms of the various 
(longitudinal and transverse) form factors. Next, we  
show  how the effective charge
and gluon mass  may be reconstructed from  the data obtained   
in a standard kinematical limit. 
In addition, we present 
a  numerical study on the propagation of the (modeled) data errors  
into the effective charge and gluon mass. 
Finally, exploiting some basic field-theoretic properties, 
we relate the conventional and BFM lattice quantity at the origin. 

\subsection{BFM on the lattice}

Within perturbation theory, the BFM was formulated to all orders on the lattice in~\cite{Luscher:1995vs}
(see also the early work of Gross and Dashen \cite{Dashen:1980vm}); however, its nonperturbative 
implementation has been pending for quite some time.
A possible nonperturbative formulation, which evades the well-known ``Neuberger 0/0 problem''~\cite{Neuberger:1986xz},
has been only recently introduced in~\cite{Binosi:2012st}, through a reformulation 
of the BFM method in terms of canonical transformations (so that dynamical ghosts are not needed).  
This led to the proposal of the gauge fixing functional~\cite{Binosi:2012st,Cucchieri:2012ii}
\be
F[g]=-\int\!\mathrm{d}^4 {x}\,\mathrm{Tr}\,(A^g_\mu-\widehat{A}_\mu)^2,
\label{BFMGFF}
\ee
which upon minimization on the group elements $g$ provides the background Landau gauge condition $\widehat{\cal D}_\mu(A^g_\mu-\widehat{A}_\mu)=0$. Then, on the minimum of this functional, the mapping given by
\be
A\to A^{g(A,\widehat{A})}-\widehat{A}\equiv Q_\mu
\label{mapping}
\ee  
defines the  action of  a canonical  transformation on  the gauge
fields, constituting a nonperturbative generalization of the familiar 
splitting into a background and a quantum field ~\cite{Abbott:1981ke}. 
The inverse of this mapping  amounts  to a  gauge  transformation,  which  can be  used  to determine the  quantum field $Q$ corresponding to  the gauge configuration minimizing  the gauge-fixing  functional~\noeq{BFMGFF}.
The relevant correlators (\eg in the two- and three-point functions) may then be constructed in terms of this 
particular field, and lattice simulations of the corresponding  discretized version can be then carried out.

Preliminary numerical simulations  for
a    variety   of   explicit    backgrounds   have    been   performed
in~\cite{Cucchieri:2012ii},    indicating that    the
functional~\noeq{BFMGFF} furnishes  a convergence rate  comparable 
to that of the  zero  background   case.  
Note, however, that 
in  the  PT-BFM
formulation  one considers  the background  field as  an  external yet
unspecified source,  to be  set to zero  after taking  the appropriate
derivatives of the vertex functional. 
Therefore, in order to properly simulate 
this procedure on the lattice, one ought to 
introduce a suitable dependence of the background  field  $\widehat{A}$ on a parameter, which, upon 
variation, would smoothly turn it off.  
Work in this direction is already  in progress; at  the moment, we
are not  aware of any  theoretical obstruction that would  prevent the computation of  PT-BFM Green's functions on the  lattice by techniques similar to those discussed in~\cite{Giusti:2001xf}.

\subsection{The basic lattice quantity}

Let us therefore assume  
that the three-point function  defined in~\1eq{3pdef} 
may be indeed simulated on the lattice, following the procedure briefly outlined above.
Then, as is customary, in the Landau gauge, one 
considers the following ratio~\cite{Cucchieri:2006tf}
\be
{\widehat R}(q,r,p)=\frac{{\cal N}(q,r,p)}{{\cal D}(q,r,p)},
 \label{Rprojection}
\ee
with the numerator and denominator given by 
\bea \label{NandD1}
 {\cal N}(q,r,p)& =& \Gamma_{\alpha\mu\nu}^{(0)}(q,r,p)
\widehat{\cal G}_{\rho\sigma\tau}(q,r,p), \nonumber \\
{\cal D}(q,r,p) &=& \Gamma_{\alpha\mu\nu}^{(0)}(q,r,p) 
P^{\alpha\rho}(q) P^{\mu\sigma}(r) P^{\nu\tau}(p) 
\Gamma_{\rho\sigma\tau}^{(0)}(q,r,p) \widehat{\Delta}_m(q)\widehat{\Delta}_m(r)\widehat{\Delta}_m(p), 
\eea
where a common color factor cancels out in the ratio. Note the index ``m'' in the full propagators, 
indicating the nonperturbative generation of a gluon mass. In addition, 
and according to the discussion in the previous section, 
the substitution given in \1eq{modifiedvertex} must also be implemented.
Then, inserting \1eq{ampdef} into ${\cal N}(q,r,p)$, together with the aforementioned substitution, 
we see that 
\n{i} the product $\widehat{\Delta}_m(q)\widehat{\Delta}_m(r)\widehat{\Delta}_m(p)$ 
cancels out when forming the ratio of \1eq{Rprojection}, and 
\n{ii} any reference to the pole vertex 
${\widehat V}_{\rho\sigma\tau}$ disappears, due to the longitudinality condition \1eq{totlon} that it satisfies.

Thus, the numerator and denominator become  
\bea
 {\cal N}(q,r,p) &=& \Gamma_{\alpha\mu\nu}^{(0)}(q,r,p) P^{\alpha}_{\rho}(q) P^{\mu}_{\sigma}(r) P^{\nu}_{\tau}(p) \widehat{\Gamma}_{m}^{\rho\sigma\tau}(q,r,p), \nonumber \\
{\cal D}(q,r,p) &=& \Gamma_{\alpha\mu\nu}^{(0)}(q,r,p) P^{\alpha\rho}(q) P^{\mu\sigma}(r) P^{\nu\tau}(p) \Gamma_{\rho\sigma\tau}^{(0)}(q,r,p).
 \label{NandD}
\eea

When one decomposes the full three-gluon vertex into a longitudinal and a transverse part, as in~\1eq{decomp}, the numerator in~\1eq{NandD} becomes
\be 
{\cal N}(q,r,p) = {\cal N}_{(\ell)}(q,r,p) + {\cal N}_{(t)}(q,r,p),
\label{numerator}
\ee
with
\be 
{\cal N}_{(\ell,t)}(q,r,p) = \Gamma_{\alpha\mu\nu}^{(0)}(q,r,p) P^\alpha_\rho(q) P^\mu_\sigma(r) P^\nu_\tau(p) 
\widehat\Gamma^{\rho\sigma\tau}_{m\,(\ell,t)}(q,r,p).
\label{ltnum}
\ee
Then, the denominator is given by
\be 
{\cal D}(q,r,p) = 4\frac{r^2 p^2 - (r\spr p)^2}{q^2 r^2 p^2} [3(q^2 r^2 + q^2 p^2 + r^2 p^2) + (r\spr p)^2 - r^2 p^2],
\label{den}
\ee  
the longitudinal part of the numerator by  
\bea
{\cal N}_{(\ell)}(q,r,p) &=& 4\frac{r^2 p^2 - (rp)^2}{q^2 r^2 p^2} \left\{ [3q^2r^2 - (q\spr p)(p\spr r)]\widehat{A}_1 + [3r^2p^2 - (p\spr q)(q\spr r)]\widehat{A}_2 \right.\nonumber \\
&+& \left.[3q^2p^2 - (q\spr r)(r\spr p)]\widehat{A}_3 + [(q\spr r)(r\spr p)(p\spr q) - q^2r^2p^2]\widehat{A}_4 \right\},
\label{longnum}
\eea
and its transverse part by 
\bea 
{\cal N}_{(t)}(q,r,p) &=& 2[r^2 p^2 - (rp)^2] \left\{ [3(q\spr r)-p^2]\widehat{Y}_1 + [3(r\spr p)-q^2]\widehat{Y}_2 \right.\nonumber \\
&+&\left. [3(q\spr p)-r^2]\widehat{Y}_3 + 6\widehat{Y}_4 \right\}. 
\label{transnum}
\eea
Note also that the identity
\be
(q\spr r)(r\spr p) + (r\spr p)(p\spr q) + (p\spr q)(q\spr r) = q^2 r^2 - (q\spr r)^2 = q^2 p^2 - (q\spr p)^2 = r^2 p^2 - (r\spr p)^2,
 \label{kineid}
\ee
which can be easily proved using momentum conservation, has been employed in deriving the above expressions, and that we have introduced the notation 
\bea
\widehat{A}_1 = \widehat{X}_1 - (q\cdot r)\widehat{X}_3; &\qquad & \qquad \widehat{A}_2 = \widehat{X}_4 - (r\spr p)\widehat{X}_6; \nonumber \\
\widehat{A}_3 = \widehat{X}_7 - (p\spr q)\widehat{X}_9; &\qquad& \qquad \widehat{A}_4 = -\widehat{X}_3 - \widehat{X}_6 - \widehat{X}_9.
\label{Afactors}
\eea

It is important to emphasize that the expressions given in \2eqs{longnum}{transnum} 
carry over directly to the case of the conventional three-gluon vertex $\Gamma_{\!m}^{\alpha\mu\nu}$, 
simply by converting the hatted quantities to normal ones. Of course, as already mentioned, 
the functional form of the corresponding $X_i$ comprising the $A_i$ is significantly more complicated.
Note also that even though 
${X}_{10}$ does not vanish, it still does not contribute to \1eq{longnum}, because 
$P^{\alpha}_{\rho}(q) P^{\mu}_{\sigma}(r) P^{\nu}_{\tau}(p) \ell_{10}^{\rho\sigma\tau} =0$.

It turns out that the physical quantity of interest may be extracted from ${\widehat R}$ by employing a  
very common kinematic choice.
Specifically, to begin with, as is customary in lattice studies, 
we will describe the ratio ${\widehat R}$ in terms of the modulo of two independent momenta 
(say, $q^2$ and $r^2$) and the angle $\phi$ formed between them; 
thus ${\widehat R} = {\widehat R}(q^2,r^2,\phi)$. 
Then, the quantity of interest corresponds to the case ${\widehat R}(q^2,0,\pi/2)$, which is 
a special case of the so-called ``orthogonal configuration'', namely ${\widehat R}(q^2,r^2,\pi/2)$.

In this latter configuration we have,
\be 
p^2 = q^2 + r^2  ; \qquad q\spr r=0 ; \qquad q\spr p=-q^2 ; \qquad r\spr p=-r^2,
\label{ortmomenta}
\ee
and, therefore, the relevant quantities reduce to
\be 
{\cal D}(q,r,\pi/2) = \frac{4}{q^2 + r^2} [3(q^4 + r^4) + 8q^2r^2],
\label{Dort}
\ee
and
\bea 
{\cal N}^{(\ell)}(q,r,\pi/2) &=& \frac{4}{q^2 + r^2}[2q^2r^2\widehat{A}_1 + 3r^2(q^2+r^2)\widehat{A}_2 + 3q^2(q^2+r^2)\widehat{A}_3 - q^2r^2(q^2+r^2)\widehat{A}_4],\nonumber\\
{\cal N}^{(t)}(q,r,\pi/2) &=& -2q^2r^2[(q^2+r^2)\widehat{Y}_1 + (3r^2+q^2)\widehat{Y}_2 + (3q^2+r^2)\widehat{Y}_3 + 6\widehat{Y}_4], 
 \label{Ntort}
\eea
with the (suppressed) arguments of the form-factors $\widehat{A}_i$ and $\widehat{Y}_i$ correspondingly adapted to the particular kinematic configuration chosen.

At this point, if, in addition, we set $r^2=0$, then {\it the transverse term vanishes}, ${\cal N}^{(t)}(q,0,\pi/2) =0$, and we obtain 
\be 
{\widehat R}(q^2,0,\pi/2) = \widehat{A}_3(q,0,\pi/2) = \widehat{X}_7 + q^2 \widehat{X}_9,
\label{Rcomlong}
\ee
so that (we only indicate the $q^2$ in the argument of ${\widehat R}$)
\begin{equation}
\widehat{R}(q^2) = [q^2 \widehat{J}_m(q^2)]^{\,\prime},
\label{hatR}
\end{equation}
where, as before, the prime indicates derivatives with respect to $q^2$.
Let us point out that this  
particular result may be derived directly from  \1eq{Rprojection}, by substituting 
in it the expression for $\widehat\Gamma_{\alpha\mu\nu}$ given in \1eq{vertexrzero}. Specifically, 
\be
{\widehat R}(q,r,p) = \widehat{J}(q^2) - 2q^2\widehat{J}'(q^2)
\left\{\frac{\Gamma_{\alpha\mu\nu}^{(0)}(q,r,p) P^{\alpha\tau}(q) P^{\mu}_{\sigma}(r) P^{\nu}_{\tau}(p) q^\sigma}{{\cal D}(q,r,p)}
\right\}.
\label{Rshort}
\ee
In the 
orthogonal configuration ($q\spr r=0$), 
we have that $q^\sigma P^{\mu}_{\sigma}(r) = q^\mu$; then, using \1eq{Dort} and setting $r^2=0$, 
it is easy to show that the quantity in curly brackets is equal to $\{-\frac{1}{2}\}$.

Then, simple integration of \1eq{hatR}  yields 
\be
q^2 \widehat{J}_m(q^2) = \int_0^{q^2} \mathrm{d}p^2 \widehat{R}(p^2) +C,
\ee
and, assuming that both $\widehat{J}_m(q^2)$ and $\widehat{R}(p^2) $ are finite for all values of the momentum, we see that
the integration constant must vanish, $C=0$. Thus, finally, one obtains the relation
\be
\widehat{J}_m(q^2) = \frac{1}{q^2}\int_0^{q^2}\!\! \mathrm{d}p^2 \,\widehat{R}(p^2).
\label{Jhat}
\ee

Let us next renormalize this result within the MOM scheme, denoting the final 
answer by $\widehat{J}_{m}^{(r)}(q^2)$. If we impose the standard MOM condition 
$\widehat{J}_{m}^{(r)}(\mu^2)=1$, at some arbitrary momentum scale $\mu$, 
then, we have that 
\be
\widehat{J}_{m}^{(r)}(q^2) = \frac{\widehat{J}_m(q^2)}{\widehat{J}_m(\mu^2)}.
\ee
However, let us point out that, strictly speaking,
due to the presence of the gluon mass, 
this last normalization condition imposed on $\widehat{J}_m$ 
cannot be enforced simultaneously with the corresponding MOM 
condition for ${\widehat\Delta}_m^{-1}(q^2)$, namely ${\widehat\Delta}_m^{-1}(\mu^2)=1$.
Indeed, since, for any arbitrary $\mu$, 
${\widehat\Delta}_m^{-1}(\mu^2)=\mu^2 \widehat{J}_m (\mu^2) + {\widehat m^2}(\mu^2) $, 
if at a given $\mu$ we impose that 
$\widehat{J}_{m}(\mu^2)=1$, then, automatically, at the same $\mu$,  
 ${\widehat\Delta}_m^{-1}(\mu^2) =\mu^2 [1 + {\widehat m^2}(\mu^2)/\mu^2]$. 
Note however, that unless one chooses to 
push the value of  $\mu$ very deep in the infrared, this discrepancy is numerically immaterial;
for example, when one renormalizes $\widehat\Delta_m$ at $\mu=4.3$ GeV, the 
value of the gluon mass at the origin is ${\widehat m}(0)=1$ GeV, making the ratio 
${\widehat m}^2(0)/\mu^2$ of the order of $5\%$. Of course, this estimate is 
just an upper bound for the relevant ratio ${m^2}(\mu^2)/\mu^2$, 
which in reality is significantly smaller, since the function 
${\widehat m^2}(q^2)$ is decreasing rather rapidly (see, \eg the inset of the left panel of \fig{startconfig}); 
in fact, at $\mu=4.3$ GeV,  the gluon mass is practically negligible.
  
\subsection{Modelling the error propagation}

In order to understand how a possible lattice signal 
for the quantity $\widehat{R}(q^2)$ of~\1eq{hatR} may provide direct information on the effective charge ${\overline\alpha}(q^2)$ 
(and the possible caveats associated with such a determination), we perform in what follows a detailed numerical study. 
Specifically, starting from the knowledge of the conventional quenched lattice propagator $\Delta_m(q^2)$ 
and ghost dressing function $F(q^2)$~\cite{Bogolubsky:2009dc}, together with the associated solution of 
the mass equation $m^2(q^2)$~\cite{Binosi:2012sj}, one can reconstruct first $\widehat{J}(q^2)$, 
and next obtain the expected shape of $\widehat{R}(q^2)$ through 
\be
\widehat{R}(q^2)=\left[\frac{\Delta^{-1}_m(q^2)-m^2(q^2)}{F^2(q^2)} \right]',
\ee
where we have used \1eq{massive} in Euclidean space. 

\begin{figure}[!t]  
\mbox{}\hspace{-1.5cm}
\includegraphics[scale=0.97]{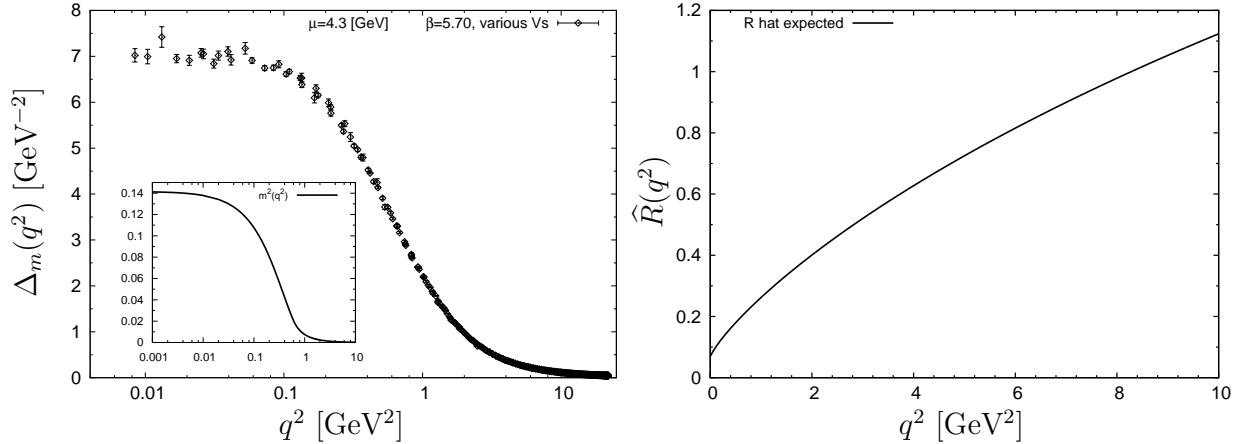}
\caption{The expected shape of the ratio $\widehat{R}(q^2)$ (right panel), obtained starting from the quenched lattice data 
for the $SU(3)$ gluon propagator $\Delta_m(q^2)$ (left panel), and the corresponding solution of the mass equation $m^2(q^2)$ (left panel, inset).}
\label{startconfig}
\end{figure}

The resulting curve (shown in \fig{startconfig}), which will be referred to as the `expected' result, can be parametrized to a high precision by the function 
\be
\widehat{R}(q^2)=A_2+ \frac{A_1-A_2}{1+\left(q^2/ q_0^2\right)^x},
\label{fit}
\ee 
with best fit parameters corresponding to the values
\be
\bar A_1 = 0.083;\qquad \bar A_2= 5.150;\qquad \bar q_0=7.156\ \mathrm{GeV};\qquad  \bar x=0.836.
\label{best-fit}
\ee


\begin{figure}[!t]
\mbox{}\hspace{-0.7cm}
\includegraphics{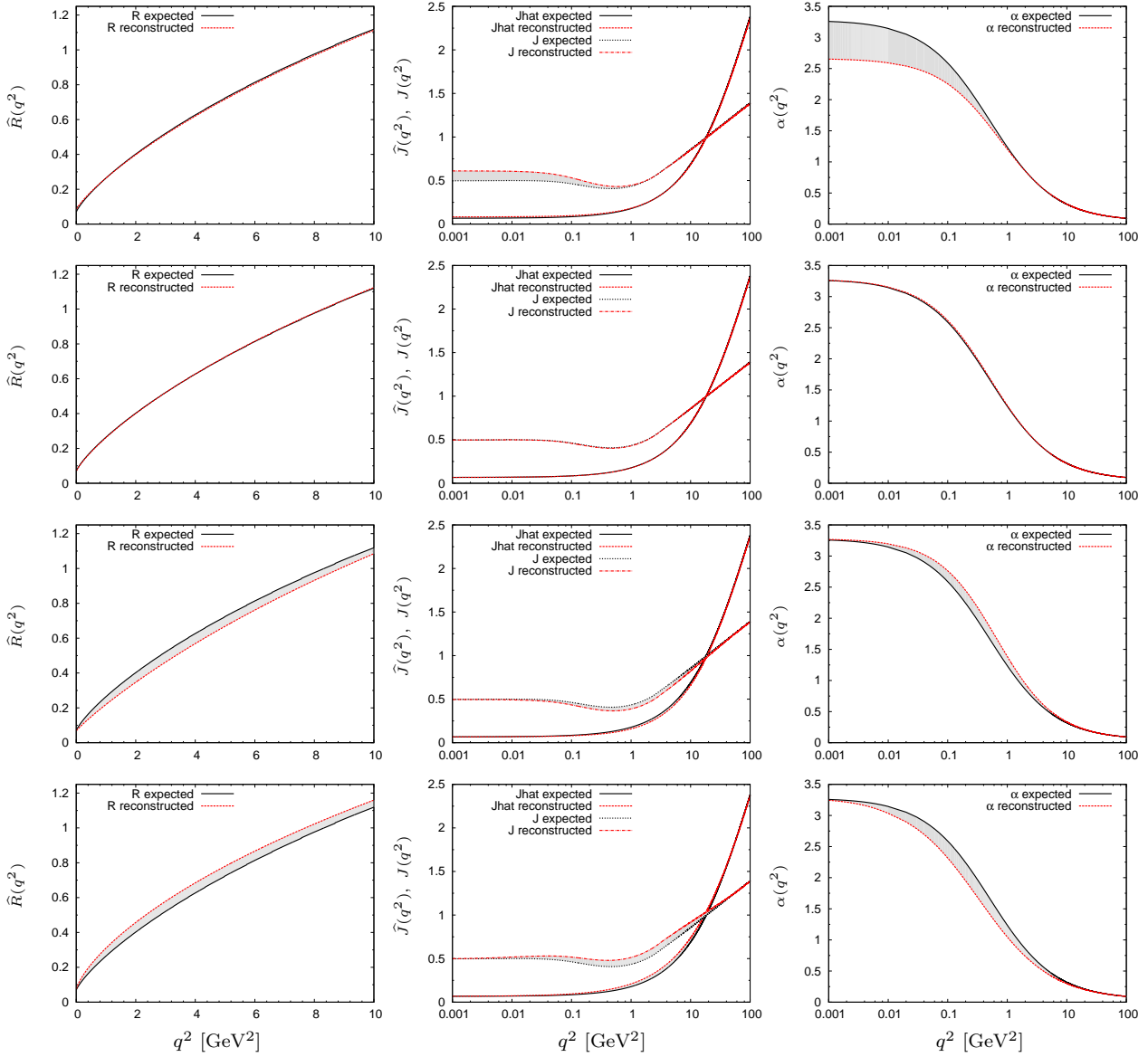} 
\caption{\label{total} (color online) Numerical study of how uncertainties in the determination of $\widehat{R}(q^2)$, 
parametrized according to~\1eq{fit}, reflect into the determination of $\widehat{J}$, $J$ as well as the effective charge ${\overline\alpha}(q^2)$. 
In particular, we compare the `expected' results with the results reconstructed from the parametrization~\noeq{fit} for the following parameter values: (first row) 
the best fit parameters~\noeq{best-fit}; (second row) $A_1$ fixed at the expected value $\widehat{R}(0)=0.0562$, and, for the remaining coefficients the refitted values $A_2=5.246$, $ q_0=7.279$ and $x=0.8167$; (third row) $A_1$ and $x$ fixed at the values $\widehat{R}(0)$ and $0.92$ respectively, and $A_2=4.632$, $ q_0= 6.234$ GeV ; (fourth row) $A_1$ and $x$ fixed at the values $\widehat{R}(0)$ and $x=0.72$ respectively, and, finally, $A_2=6.251$ $ q_0=9.21$ GeV.} 
\end{figure} 


The above functional form of the expected behavior of $\widehat{R}(q^2)$ 
is rather useful, because it allows for a systematic analysis of how uncertainties, simulated through deviations of the    
fitting parameters from their ``optimal'' values,  
can influence the reconstruction of the effective charge (and gluon mass, see below). 

The results of this study are shown in~\fig{total}. As can be seen, 
the most sizable deviation between the `expected' and the reconstructed results occurs when 
uncertainties in the determination of $\widehat{R}(0)$ are sizable (\fig{total}, first row). 
This is mainly due to the fact that the `expected' value for $\widehat{R}(0)$ turns out to be of ${\cal O}(10^{-2})$, 
and moderate deviations imply a considerable effect in the determination of $\widehat{J}_m$ through~\1eq{Jhat}; this, 
in turn, translates into a large variation of 
the effective charge, given  that ${\overline\alpha}\sim\widehat{J}_m^{-1}$, see~\1eq{effch}. 

Specifically, in the first row of~\fig{total} we show the results for $\widehat{R}$, $\widehat{J}$, $J$ and ${\overline\alpha}$ 
obtained starting from~\1eq{fit} with the best fit parameters~\noeq{best-fit}; since in the parametrization~\noeq{fit}$, \widehat{R}(0)\equiv\bar A_1$, 
such fit overestimates the value at the origin; this error propagates 
in the determination of a reconstructed effective charge, which comes out 
suppressed in the IR with respect to the `expected' value.

The following three rows in \fig{total} show the effect of 
uncertainties in the determination of the exponent $x$ (which controls the overall shape of the $\widehat{R}$ curve),  
once the behavior at the origin has been fixed to its `expected' value, that is $A_1$ is fixed to the value $\widehat{R}(0)$. 
As can be seen, the effect is significantly milder, and one can always reconstruct the effective charge to a high degree of accuracy.

\begin{figure}
\includegraphics[scale=0.6]{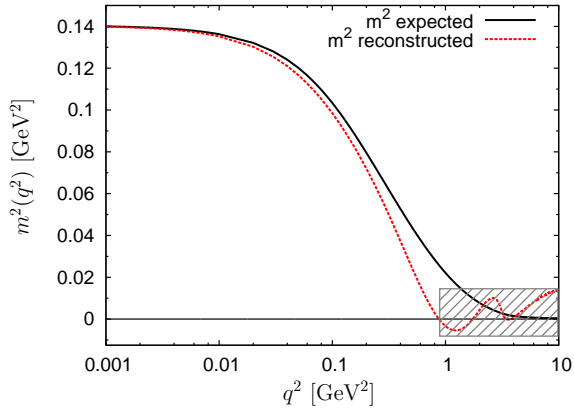}
\caption{\label{mass}The reconstructed running mass starting from $\widehat{R}(q^2)$ 
as given in \1eq{fit} when the parameters chosen are the best fit ones of \1eq{best-fit}. Notice the problems that afflict the determination of the UV mass tail (shaded region).}
\end{figure}  

Finally, let us focus our attention to the mass $m^2(q^2)$; as we will see, 
its reconstruction from this particular type of lattice measurements is especially subtle. 
The main difficulty stems from the fact that the extraction of $m^2(q^2)$ proceeds by means of the relation 
\be
m^2(q^2)= \Delta^{-1}_m(q^2)-q^2J_m(q^2).
\label{emsquared}
\ee
Now, whereas both terms on the rhs increase in the UV, on theoretical grounds we know that $m^2(q^2)$ must decrease rather rapidly;
this, in turn, implies that a rather delicate cancellation between these aforementioned two terms must take place. 
This cancellation, however, is very likely to be distorted by the 
reconstruction procedure, especially for high values of $q^2$. This particular problem is shown in \fig{mass}, where we plot 
the reconstructed mass for $\widehat{R}(q^2)$ provided by \2eqs{fit}{best-fit}. As can be seen there, 
while the determination in the IR is reasonably accurate, the tail is seriously distorted, displaying even negative regions. 
This characteristic pathology persists (with various degrees of intensity) in all cases analyzed in~\fig{total}.
 
\subsection{Relating $\widehat{R}(0)$ with $R(0)$}

Let us now consider the conventional ${R}(q^2)$, obtained from \1eq{Rcomlong} 
through the direct substitution ${\widehat X}_{7,9}\to {X}_{7,9}$. Obviously, the 
complicated structure of ${X}_{7}$ and  ${X}_{9}$ [see, e.g., \1eq{X7}], infested by the 
unknown form factors of the ghost-gluon kernel, makes their use for arbitrary $q^2$
impractical. However, when $q^2=0$ the corresponding expressions simplify 
substantially, providing a fairly simple expression for ${R}(0)$.  

Specifically,  after implementing ${\widehat X}_{7,9}\to {X}_{7,9}$ in Eq.~(\ref{Rcomlong}), let us set $q^2=0$, to obtain 
\begin{equation}\label{Rzero}
R(0) = F(0)J_m(0)a(0,0,0),
\end{equation}
where Eq.~(\ref{X7}) has been employed.


\begin{figure}[t]
\includegraphics[scale=0.65]{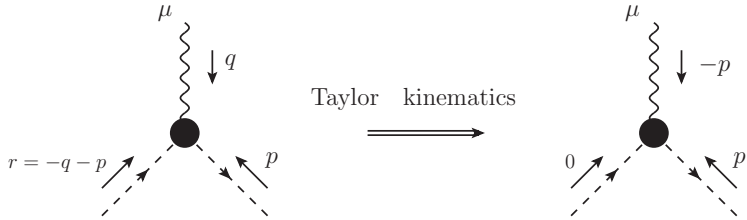}
\caption{The ghost-gluon vertex and the Taylor kinematics.}
\label{Taylor}
\end{figure}


At this point one may invoke Taylor's theorem~\cite{Taylor:1971ff} in order to determine, under mild assumptions, the value of $a(0,0,0)$.
To that end, consider 
the general Lorentz decomposition for the ghost-gluon vertex, 
\begin{equation}\label{glghvertex}
\Gamma_\mu(r,q,p) = B_1(r,q,p)q_\mu - B_2(r,q,p)p_\mu.
\end{equation}
Notice that, setting $B_1^{(0)}=0$ and $B_2^{(0)}=1$, one recovers the tree-level value of the ghost-gluon vertex $\Gamma_\mu^{(0)}=-p_\mu$. In the Taylor kinematic configuration, see Fig.~\ref{Taylor}, Eq.~(\ref{glghvertex}) becomes
\begin{equation}\label{glghTaylor}
\Gamma_\mu(0,-p,p) = -[B_1(0,-p,p) + B_2(0,-p,p)]p_\mu.
\end{equation}
Taylor's theorem states that 
\begin{equation}\label{linkTaylor}
B_1(0,-p,p) + B_2(0,-p,p) = 1,
\end{equation}
to all-orders in perturbation theory.

On the other hand, it is well-known that $\Gamma_\mu(r,q,p)$ can be obtained from the contraction
\begin{eqnarray}
\Gamma_\mu(r,q,p) &=& -p^\nu H_{\nu\mu}(p,r,q) \nonumber \\ 
&=& -[a_{qrp} + (q\spr p) b_{qrp} + (q\spr p) d_{qrp} + p^2 e_{qrp}]p_\mu - [(q\spr p)b_{qrp} + p^2 c_{qrp}]q_\mu,\hspace{1cm}
\label{gluonghostH}
\end{eqnarray}
which, in the Taylor kinematics reduces to
\begin{equation}\label{gammazero}
\Gamma_\mu(0,-p,p) = -\lbrace a(-p,0,p) - p^2[c(-p,0,p) + d(-p,0,p) - e(-p,0,p)]\rbrace p_\mu.
\end{equation}
Thus, equating Eq.~(\ref{glghTaylor}) with Eq.~(\ref{gammazero}) and using Eq.~(\ref{linkTaylor}), we deduce the constraint
\begin{equation}
\label{constTaylor}
a(-p,0,p) - p^2[c(-p,0,p) + d(-p,0,p) - e(-p,0,p)] = 1.
\end{equation}

Finally, taking the limit $p\rightarrow 0$ in Eq.~(\ref{constTaylor}), 
and assuming that $c$,$d$, and $e$ are regular functions in that limit 
one obtains that $a(0,0,0)$ assumes its tree-level value, 
\begin{equation}\label{specialvalue}
a(0,0,0) = 1 .
\end{equation}
Consequently, \1eq{Rzero} becomes 
\begin{equation}\label{Rzero2}
R(0) = F(0)J_m(0) ,
\end{equation}
whereas, in the same limit, Eq.~(\ref{hatR}) gives directly
\begin{equation}\label{Rhatzero}
\widehat{R}(0) = \widehat{J}_m(0).
\end{equation}

It is now relatively straightforward to derive a simple relation relating $R(0)$ with $\widehat{R}(0)$. 
Specifically, use of \1eq{BQIJ} and the relation~\noeq{funrel} at $q^2=0$, yields
\begin{equation}\label{BQIJzer}
\widehat{J}_m(0) = [1+G(0)]^2 J_m(0)=F^{-2}(0)J_m(0),
\end{equation}
so that finally 
\begin{equation}
\frac{R(0)}{\widehat{R}(0)} = F^3(0).
\label{quotientR}
\end{equation}

Since $F(0)>1$, the determination of $R(0)$ from $\widehat{R}(0)$ might be error prone; in the  case studied in the previous section, at $\mu=4.3$ GeV one has $F(0)=2.86$, which furnishes $R(0)= 1.58$ for the initial value $\widehat{R}(0)=0.067$. Therefore, given also the sensitivity of the effective charge to the value of $\widehat{R}(0)$, one should ideally proceed the other way round, limiting the possible values of  $\widehat{R}(0)$ through an independent  measurement of $R(0)$.

\section{Conclusions}\label{sec:conc}

In this work we have explored the possibility of determining the complete momentum evolution of the 
QCD effective charge from a special kinematic limit of the 
three-gluon vertex corresponding to three background gluons. 
Given that within the  BFM quantization scheme the (background) gauge invariance is preserved,
the aforementioned vertex satisfies linear WIs, with no reference to the ghost sector;
within the PT 
the same property emerges naturally, after the systematic rearrangement 
of an appropriate observable, following the standard pinching rules. 
Consequently, and in contradistinction to what happens in the case of the conventional 
three-gluon vertex, the 
longitudinal form factors of this PT-BFM vertex may be expressed {\it exclusively} in terms of 
the background gluon wave function $\widehat{J}_m$.
By virtue of the dynamically generated gluon mass, this latter quantity, 
as well as the physical effective charge defined from it,  
are infrared finite and free of any 
divergences related to the perturbative Landau pole.

Particularly interesting in this context is the possibility of 
simulating the (Landau gauge) PT-BFM vertex on the lattice. To that end, 
after briefly reviewing the general theoretical feasibility of such a task,   
we have focused on the specifics of the vertex simulation, 
with special emphasis 
on the relevant kinematic limit that projects out 
the desired quantity $\widehat{J}_m(q^2)$. 
In addition, a preliminary numerical analysis suggests that the 
extraction of the effective charge is relatively 
insensitive to the numerical uncertainties that may infest an actual lattice 
simulation. On the contrary,  
the reconstruction of the running gluon mass turned out to be considerably more subtle; 
specifically,  instead of being positive-definite and monotonically decreasing, 
the corresponding curve displays unphysical fluctuations past $1$ GeV, 
due to the distortion of delicate numerical cancellations.

The results presented in this article are expected to contribute to 
the collective effort dedicated to the deeper comprehension of the nature and 
properties of the QCD effective charge.
In fact, the possibility of probing directly  
the value of $\overline\alpha(q^2)$ at the origin, through the corresponding 
lattice extraction of $\widehat{J}_m(0)$, is intriguing,
and may serve as testing ground for various alternative pictures. 
In particular, 
the value of $\overline\alpha(0)$ 
is directly related to the notion of the ``QCD conformal window'', 
appearing in studies based on the AdS/QCD correspondence~\cite{Brodsky:2010ur}. 

It is important to emphasize that the nature of the observable ${\widehat R}$, coupled 
with the fact that we work in the Landau gauge, results in  
the total annihilation of the completely longitudinal pole vertex ${\widehat V}$, which is 
intimately associated with the Schwinger mechanism of gauge-boson mass generation. 
In that sense, the situation is completely analogous to
what happens with standard observables, where all direct effects  
from this particular vertex vanish, due to  current conservation or general on-shellness conditions.
Note that the same situation applies to the case of conventional three-gluon vertex, 
and its associated $V$;
they too cancel out completely from the corresponding lattice quantity $R$.
Therefore, any (apparently) singular behavior that may be observed in simulations of these quantities 
should not be interpreted as a potential consequence of the pole vertices. 

A possible determination of $\widehat{J}_m(q^2)$ can provide considerable 
theoretical insights that extend beyond the accurate extraction 
of the effective charge, and could help us explore the 
nonperturbative behavior of additional key dynamical ingredients. 
In particular, 
the complete integral equation that 
govern the evolution of  $\widehat{J}_m(q^2)$ 
depends on the fully dressed four-gluon vertex that involves  
one background and three quantum gluons. The available information on this 
vertex is very limited at the moment; the only robust result 
known is the WI that is satisfies when contracted 
with the momentum of the  background gluon. 
It is therefore  reasonable to expect that 
further research will be devoted in this direction. 
Then it is clear that independent information on  $\widehat{J}_m(q^2)$ may prove valuable 
for determining 
the structure of this elusive vertex, at least in some simple kinematic limits. 

In general, the lattice simulation of the PT-BFM propagator and vertices   
would  offer the unique opportunity to verify explicitly 
powerful formal relations, 
connecting basic Green's function of the theory. 
Note, for instance, that the simulation of $\widehat{\Delta}(q^2)$
could furnish a direct confirmation of the fundamental BQI of \1eq{BQIpropagatorsa}, given that both 
${\Delta}(q^2)$ and $G(q^2)$ have already been simulated on the lattice. 
In addition, the result of \1eq{quotientR} may be of certain usefulness for future lattice endeavors.
Specifically, one may use combinations of lattice results to probe the veracity of the (few) theoretical assumptions entering in its derivation; conversely, 
one may assume the validity of~\1eq{quotientR} and use it to validate the lattice implementation of the BFM algorithm. To be sure, such comparisons must be carried out between lattice simulations possessing similar  parameters (\eg bare couplings, spacings,~volumes).  

Finally, note that the present considerations may be extended to include other fundamental vertices of 
the PT-BFM formalism, such as the vertex connecting a background gluon and a 
ghost-anti-ghost pair, usually denoted by $\widehat{\Gamma}_{\mu}$. 
This vertex has a rather reduced tensorial structure, and 
satisfies a simple Abelian WI, relating its divergence to the difference of two 
inverse ghost propagators. 
A preliminary study reveals that relations similar to~\noeq{vertexrzero}
may be also obtained for the ghost dressing function. 
Hence, lattice simulation of  $\widehat{\Gamma}_{\mu}$ may 
provide nontrivial cross-checks on the infrared behavior of this latter quantity. 
We hope to present the full details of the related analysis in the near future. 
 
\acknowledgments

The research of D. I. and J. P. is supported by the Spanish MEYC under Grant No. FPA2011-23596.

\bibliography{../../../../Bibliography/bibliography}

\begin{thebibliography}{53}
\expandafter\ifx\csname natexlab\endcsname\relax\def\natexlab#1{#1}\fi
\expandafter\ifx\csname bibnamefont\endcsname\relax
  \def\bibnamefont#1{#1}\fi
\expandafter\ifx\csname bibfnamefont\endcsname\relax
  \def\bibfnamefont#1{#1}\fi
\expandafter\ifx\csname citenamefont\endcsname\relax
  \def\citenamefont#1{#1}\fi
\expandafter\ifx\csname url\endcsname\relax
  \def\url#1{\texttt{#1}}\fi
\expandafter\ifx\csname urlprefix\endcsname\relax\def\urlprefix{URL }\fi
\providecommand{\bibinfo}[2]{#2}
\providecommand{\eprint}[2][]{\url{#2}}

\bibitem[{\citenamefont{Cornwall}(1982)}]{Cornwall:1981zr}
\bibinfo{author}{\bibfnamefont{J.~M.} \bibnamefont{Cornwall}},
  \bibinfo{journal}{Phys. Rev.} \textbf{\bibinfo{volume}{D26}},
  \bibinfo{pages}{1453} (\bibinfo{year}{1982}).

\bibitem[{\citenamefont{Cucchieri and Mendes}(2007)}]{Cucchieri:2007md}
\bibinfo{author}{\bibfnamefont{A.}~\bibnamefont{Cucchieri}} \bibnamefont{and}
  \bibinfo{author}{\bibfnamefont{T.}~\bibnamefont{Mendes}}
  (\bibinfo{year}{2007}), \eprint{arXiv:0710.0412 [hep-lat]}.

\bibitem[{\citenamefont{Bogolubsky et~al.}(2009)\citenamefont{Bogolubsky,
  Ilgenfritz, Muller-Preussker, and Sternbeck}}]{Bogolubsky:2009dc}
\bibinfo{author}{\bibfnamefont{I.}~\bibnamefont{Bogolubsky}},
  \bibinfo{author}{\bibfnamefont{E.}~\bibnamefont{Ilgenfritz}},
  \bibinfo{author}{\bibfnamefont{M.}~\bibnamefont{Muller-Preussker}},
  \bibnamefont{and}
  \bibinfo{author}{\bibfnamefont{A.}~\bibnamefont{Sternbeck}},
  \bibinfo{journal}{Phys.Lett.} \textbf{\bibinfo{volume}{B676}},
  \bibinfo{pages}{69} (\bibinfo{year}{2009}).

\bibitem[{\citenamefont{Aguilar et~al.}(2008)\citenamefont{Aguilar, Binosi, and
  Papavassiliou}}]{Aguilar:2008fh}
\bibinfo{author}{\bibfnamefont{A.}~\bibnamefont{Aguilar}},
  \bibinfo{author}{\bibfnamefont{D.}~\bibnamefont{Binosi}}, \bibnamefont{and}
  \bibinfo{author}{\bibfnamefont{J.}~\bibnamefont{Papavassiliou}},
  \bibinfo{journal}{PoS} \textbf{\bibinfo{volume}{LC2008}},
  \bibinfo{pages}{050} (\bibinfo{year}{2008}).

\bibitem[{\citenamefont{Aguilar
  et~al.}(2009{\natexlab{a}})\citenamefont{Aguilar, Binosi, Papavassiliou, and
  Rodriguez-Quintero}}]{Aguilar:2009nf}
\bibinfo{author}{\bibfnamefont{A.}~\bibnamefont{Aguilar}},
  \bibinfo{author}{\bibfnamefont{D.}~\bibnamefont{Binosi}},
  \bibinfo{author}{\bibfnamefont{J.}~\bibnamefont{Papavassiliou}},
  \bibnamefont{and}
  \bibinfo{author}{\bibfnamefont{J.}~\bibnamefont{Rodriguez-Quintero}},
  \bibinfo{journal}{Phys.Rev.} \textbf{\bibinfo{volume}{D80}},
  \bibinfo{pages}{085018} (\bibinfo{year}{2009}{\natexlab{a}}).

\bibitem[{\citenamefont{Aguilar et~al.}(2010)\citenamefont{Aguilar, Binosi, and
  Papavassiliou}}]{Aguilar:2010gm}
\bibinfo{author}{\bibfnamefont{A.}~\bibnamefont{Aguilar}},
  \bibinfo{author}{\bibfnamefont{D.}~\bibnamefont{Binosi}}, \bibnamefont{and}
  \bibinfo{author}{\bibfnamefont{J.}~\bibnamefont{Papavassiliou}},
  \bibinfo{journal}{JHEP} \textbf{\bibinfo{volume}{1007}}, \bibinfo{pages}{002}
  (\bibinfo{year}{2010}).

\bibitem[{\citenamefont{Dokshitzer et~al.}(1996)\citenamefont{Dokshitzer,
  Marchesini, and Webber}}]{Dokshitzer:1995qm}
\bibinfo{author}{\bibfnamefont{Y.~L.} \bibnamefont{Dokshitzer}},
  \bibinfo{author}{\bibfnamefont{G.}~\bibnamefont{Marchesini}},
  \bibnamefont{and} \bibinfo{author}{\bibfnamefont{B.~R.}
  \bibnamefont{Webber}}, \bibinfo{journal}{Nucl. Phys.}
  \textbf{\bibinfo{volume}{B469}}, \bibinfo{pages}{93} (\bibinfo{year}{1996}).

\bibitem[{\citenamefont{von Smekal et~al.}(1997)\citenamefont{von Smekal,
  Alkofer, and Hauck}}]{vonSmekal:1997is}
\bibinfo{author}{\bibfnamefont{L.}~\bibnamefont{von Smekal}},
  \bibinfo{author}{\bibfnamefont{R.}~\bibnamefont{Alkofer}}, \bibnamefont{and}
  \bibinfo{author}{\bibfnamefont{A.}~\bibnamefont{Hauck}},
  \bibinfo{journal}{Phys. Rev. Lett.} \textbf{\bibinfo{volume}{79}},
  \bibinfo{pages}{3591} (\bibinfo{year}{1997}).

\bibitem[{\citenamefont{Shirkov and Solovtsov}(1997)}]{Shirkov:1997wi}
\bibinfo{author}{\bibfnamefont{D.~V.} \bibnamefont{Shirkov}} \bibnamefont{and}
  \bibinfo{author}{\bibfnamefont{I.~L.} \bibnamefont{Solovtsov}},
  \bibinfo{journal}{Phys. Rev. Lett.} \textbf{\bibinfo{volume}{79}},
  \bibinfo{pages}{1209} (\bibinfo{year}{1997}).

\bibitem[{\citenamefont{Aguilar et~al.}(2003)\citenamefont{Aguilar, Natale, and
  Rodrigues~da Silva}}]{Aguilar:2002tc}
\bibinfo{author}{\bibfnamefont{A.~C.} \bibnamefont{Aguilar}},
  \bibinfo{author}{\bibfnamefont{A.~A.} \bibnamefont{Natale}},
  \bibnamefont{and} \bibinfo{author}{\bibfnamefont{P.~S.}
  \bibnamefont{Rodrigues~da Silva}}, \bibinfo{journal}{Phys. Rev. Lett.}
  \textbf{\bibinfo{volume}{90}}, \bibinfo{pages}{152001}
  (\bibinfo{year}{2003}).

\bibitem[{\citenamefont{Prosperi et~al.}(2007)\citenamefont{Prosperi, Raciti,
  and Simolo}}]{Prosperi:2006hx}
\bibinfo{author}{\bibfnamefont{G.~M.} \bibnamefont{Prosperi}},
  \bibinfo{author}{\bibfnamefont{M.}~\bibnamefont{Raciti}}, \bibnamefont{and}
  \bibinfo{author}{\bibfnamefont{C.}~\bibnamefont{Simolo}},
  \bibinfo{journal}{Prog. Part. Nucl. Phys.} \textbf{\bibinfo{volume}{58}},
  \bibinfo{pages}{387} (\bibinfo{year}{2007}).

\bibitem[{\citenamefont{Fischer}(2006)}]{Fischer:2006ub}
\bibinfo{author}{\bibfnamefont{C.~S.} \bibnamefont{Fischer}},
  \bibinfo{journal}{J. Phys.} \textbf{\bibinfo{volume}{G32}},
  \bibinfo{pages}{R253} (\bibinfo{year}{2006}).

\bibitem[{\citenamefont{Brodsky et~al.}(2010)\citenamefont{Brodsky,
  de~Teramond, and Deur}}]{Brodsky:2010ur}
\bibinfo{author}{\bibfnamefont{S.~J.} \bibnamefont{Brodsky}},
  \bibinfo{author}{\bibfnamefont{G.~F.} \bibnamefont{de~Teramond}},
  \bibnamefont{and} \bibinfo{author}{\bibfnamefont{A.}~\bibnamefont{Deur}},
  \bibinfo{journal}{Phys.Rev.} \textbf{\bibinfo{volume}{D81}},
  \bibinfo{pages}{096010} (\bibinfo{year}{2010}).

\bibitem[{\citenamefont{Courtoy and Liuti}(2013)}]{Courtoy:2013qca}
\bibinfo{author}{\bibfnamefont{A.}~\bibnamefont{Courtoy}} \bibnamefont{and}
  \bibinfo{author}{\bibfnamefont{S.}~\bibnamefont{Liuti}}
  (\bibinfo{year}{2013}).

\bibitem[{\citenamefont{Cornwall and Papavassiliou}(1989)}]{Cornwall:1989gv}
\bibinfo{author}{\bibfnamefont{J.~M.} \bibnamefont{Cornwall}} \bibnamefont{and}
  \bibinfo{author}{\bibfnamefont{J.}~\bibnamefont{Papavassiliou}},
  \bibinfo{journal}{Phys. Rev.} \textbf{\bibinfo{volume}{D40}},
  \bibinfo{pages}{3474} (\bibinfo{year}{1989}).

\bibitem[{\citenamefont{Pilaftsis}(1997)}]{Pilaftsis:1996fh}
\bibinfo{author}{\bibfnamefont{A.}~\bibnamefont{Pilaftsis}},
  \bibinfo{journal}{Nucl. Phys.} \textbf{\bibinfo{volume}{B487}},
  \bibinfo{pages}{467} (\bibinfo{year}{1997}).

\bibitem[{\citenamefont{Binosi and
  Papavassiliou}(2002{\natexlab{a}})}]{Binosi:2002ft}
\bibinfo{author}{\bibfnamefont{D.}~\bibnamefont{Binosi}} \bibnamefont{and}
  \bibinfo{author}{\bibfnamefont{J.}~\bibnamefont{Papavassiliou}},
  \bibinfo{journal}{Phys. Rev.} \textbf{\bibinfo{volume}{D66}},
  \bibinfo{pages}{111901(R)} (\bibinfo{year}{2002}{\natexlab{a}}).

\bibitem[{\citenamefont{Binosi and Papavassiliou}(2004)}]{Binosi:2003rr}
\bibinfo{author}{\bibfnamefont{D.}~\bibnamefont{Binosi}} \bibnamefont{and}
  \bibinfo{author}{\bibfnamefont{J.}~\bibnamefont{Papavassiliou}},
  \bibinfo{journal}{J.Phys.G} \textbf{\bibinfo{volume}{G30}},
  \bibinfo{pages}{203} (\bibinfo{year}{2004}).

\bibitem[{\citenamefont{Binosi and Papavassiliou}(2009)}]{Binosi:2009qm}
\bibinfo{author}{\bibfnamefont{D.}~\bibnamefont{Binosi}} \bibnamefont{and}
  \bibinfo{author}{\bibfnamefont{J.}~\bibnamefont{Papavassiliou}},
  \bibinfo{journal}{Phys.Rept.} \textbf{\bibinfo{volume}{479}},
  \bibinfo{pages}{1} (\bibinfo{year}{2009}), \bibinfo{note}{245 pages, 92
  figures}.

\bibitem[{\citenamefont{Abbott}(1981)}]{Abbott:1980hw}
\bibinfo{author}{\bibfnamefont{L.~F.} \bibnamefont{Abbott}},
  \bibinfo{journal}{Nucl. Phys.} \textbf{\bibinfo{volume}{B185}},
  \bibinfo{pages}{189} (\bibinfo{year}{1981}).

\bibitem[{\citenamefont{Aguilar and Papavassiliou}(2006)}]{Aguilar:2006gr}
\bibinfo{author}{\bibfnamefont{A.~C.} \bibnamefont{Aguilar}} \bibnamefont{and}
  \bibinfo{author}{\bibfnamefont{J.}~\bibnamefont{Papavassiliou}},
  \bibinfo{journal}{JHEP} \textbf{\bibinfo{volume}{12}}, \bibinfo{pages}{012}
  (\bibinfo{year}{2006}).

\bibitem[{\citenamefont{Binosi and
  Papavassiliou}(2008{\natexlab{a}})}]{Binosi:2007pi}
\bibinfo{author}{\bibfnamefont{D.}~\bibnamefont{Binosi}} \bibnamefont{and}
  \bibinfo{author}{\bibfnamefont{J.}~\bibnamefont{Papavassiliou}},
  \bibinfo{journal}{Phys.Rev.} \textbf{\bibinfo{volume}{D77}},
  \bibinfo{pages}{061702} (\bibinfo{year}{2008}{\natexlab{a}}).

\bibitem[{\citenamefont{Binosi and
  Papavassiliou}(2008{\natexlab{b}})}]{Binosi:2008qk}
\bibinfo{author}{\bibfnamefont{D.}~\bibnamefont{Binosi}} \bibnamefont{and}
  \bibinfo{author}{\bibfnamefont{J.}~\bibnamefont{Papavassiliou}},
  \bibinfo{journal}{JHEP} \textbf{\bibinfo{volume}{0811}}, \bibinfo{pages}{063}
  (\bibinfo{year}{2008}{\natexlab{b}}).

\bibitem[{\citenamefont{Grassi et~al.}(2001)\citenamefont{Grassi, Hurth, and
  Steinhauser}}]{Grassi:1999tp}
\bibinfo{author}{\bibfnamefont{P.~A.} \bibnamefont{Grassi}},
  \bibinfo{author}{\bibfnamefont{T.}~\bibnamefont{Hurth}}, \bibnamefont{and}
  \bibinfo{author}{\bibfnamefont{M.}~\bibnamefont{Steinhauser}},
  \bibinfo{journal}{Annals Phys.} \textbf{\bibinfo{volume}{288}},
  \bibinfo{pages}{197} (\bibinfo{year}{2001}).

\bibitem[{\citenamefont{Binosi and
  Papavassiliou}(2002{\natexlab{b}})}]{Binosi:2002ez}
\bibinfo{author}{\bibfnamefont{D.}~\bibnamefont{Binosi}} \bibnamefont{and}
  \bibinfo{author}{\bibfnamefont{J.}~\bibnamefont{Papavassiliou}},
  \bibinfo{journal}{Phys.Rev.} \textbf{\bibinfo{volume}{D66}},
  \bibinfo{pages}{025024} (\bibinfo{year}{2002}{\natexlab{b}}).

\bibitem[{\citenamefont{Grassi et~al.}(2004)\citenamefont{Grassi, Hurth, and
  Quadri}}]{Grassi:2004yq}
\bibinfo{author}{\bibfnamefont{P.~A.} \bibnamefont{Grassi}},
  \bibinfo{author}{\bibfnamefont{T.}~\bibnamefont{Hurth}}, \bibnamefont{and}
  \bibinfo{author}{\bibfnamefont{A.}~\bibnamefont{Quadri}},
  \bibinfo{journal}{Phys. Rev.} \textbf{\bibinfo{volume}{D70}},
  \bibinfo{pages}{105014} (\bibinfo{year}{2004}).

\bibitem[{\citenamefont{Aguilar
  et~al.}(2009{\natexlab{b}})\citenamefont{Aguilar, Binosi, and
  Papavassiliou}}]{Aguilar:2009pp}
\bibinfo{author}{\bibfnamefont{A.}~\bibnamefont{Aguilar}},
  \bibinfo{author}{\bibfnamefont{D.}~\bibnamefont{Binosi}}, \bibnamefont{and}
  \bibinfo{author}{\bibfnamefont{J.}~\bibnamefont{Papavassiliou}},
  \bibinfo{journal}{JHEP} \textbf{\bibinfo{volume}{0911}}, \bibinfo{pages}{066}
  (\bibinfo{year}{2009}{\natexlab{b}}).

\bibitem[{\citenamefont{Aguilar and Papavassiliou}(2010)}]{Aguilar:2009ke}
\bibinfo{author}{\bibfnamefont{A.~C.} \bibnamefont{Aguilar}} \bibnamefont{and}
  \bibinfo{author}{\bibfnamefont{J.}~\bibnamefont{Papavassiliou}},
  \bibinfo{journal}{Phys.Rev.} \textbf{\bibinfo{volume}{D81}},
  \bibinfo{pages}{034003} (\bibinfo{year}{2010}).

\bibitem[{\citenamefont{Aguilar et~al.}(2011)\citenamefont{Aguilar, Binosi, and
  Papavassiliou}}]{Aguilar:2011ux}
\bibinfo{author}{\bibfnamefont{A.}~\bibnamefont{Aguilar}},
  \bibinfo{author}{\bibfnamefont{D.}~\bibnamefont{Binosi}}, \bibnamefont{and}
  \bibinfo{author}{\bibfnamefont{J.}~\bibnamefont{Papavassiliou}},
  \bibinfo{journal}{Phys.Rev.} \textbf{\bibinfo{volume}{D84}},
  \bibinfo{pages}{085026} (\bibinfo{year}{2011}).

\bibitem[{\citenamefont{Binosi et~al.}(2012)\citenamefont{Binosi, Ibanez, and
  Papavassiliou}}]{Binosi:2012sj}
\bibinfo{author}{\bibfnamefont{D.}~\bibnamefont{Binosi}},
  \bibinfo{author}{\bibfnamefont{D.}~\bibnamefont{Ibanez}}, \bibnamefont{and}
  \bibinfo{author}{\bibfnamefont{J.}~\bibnamefont{Papavassiliou}},
  \bibinfo{journal}{Phys.Rev.} \textbf{\bibinfo{volume}{D86}},
  \bibinfo{pages}{085033} (\bibinfo{year}{2012}).

\bibitem[{\citenamefont{Binger and Brodsky}(2006)}]{Binger:2006sj}
\bibinfo{author}{\bibfnamefont{M.}~\bibnamefont{Binger}} \bibnamefont{and}
  \bibinfo{author}{\bibfnamefont{S.~J.} \bibnamefont{Brodsky}},
  \bibinfo{journal}{Phys. Rev.} \textbf{\bibinfo{volume}{D74}},
  \bibinfo{pages}{054016} (\bibinfo{year}{2006}).

\bibitem[{\citenamefont{Binosi and Quadri}(2012)}]{Binosi:2012st}
\bibinfo{author}{\bibfnamefont{D.}~\bibnamefont{Binosi}} \bibnamefont{and}
  \bibinfo{author}{\bibfnamefont{A.}~\bibnamefont{Quadri}},
  \bibinfo{journal}{Phys.Rev.} \textbf{\bibinfo{volume}{D85}},
  \bibinfo{pages}{121702} (\bibinfo{year}{2012}).

\bibitem[{\citenamefont{Cucchieri and Mendes}(2012)}]{Cucchieri:2012ii}
\bibinfo{author}{\bibfnamefont{A.}~\bibnamefont{Cucchieri}} \bibnamefont{and}
  \bibinfo{author}{\bibfnamefont{T.}~\bibnamefont{Mendes}},
  \bibinfo{journal}{Phys.Rev.} \textbf{\bibinfo{volume}{D86}},
  \bibinfo{pages}{071503} (\bibinfo{year}{2012}).

\bibitem[{\citenamefont{Marciano and Pagels}(1978)}]{Marciano:1977su}
\bibinfo{author}{\bibfnamefont{W.~J.} \bibnamefont{Marciano}} \bibnamefont{and}
  \bibinfo{author}{\bibfnamefont{H.}~\bibnamefont{Pagels}},
  \bibinfo{journal}{Phys. Rept.} \textbf{\bibinfo{volume}{36}},
  \bibinfo{pages}{137} (\bibinfo{year}{1978}).

\bibitem[{\citenamefont{Ball and Chiu}(1980)}]{Ball:1980ax}
\bibinfo{author}{\bibfnamefont{J.~S.} \bibnamefont{Ball}} \bibnamefont{and}
  \bibinfo{author}{\bibfnamefont{T.-W.} \bibnamefont{Chiu}},
  \bibinfo{journal}{Phys. Rev.} \textbf{\bibinfo{volume}{D22}},
  \bibinfo{pages}{2550} (\bibinfo{year}{1980}).

\bibitem[{\citenamefont{Cucchieri et~al.}(2006)\citenamefont{Cucchieri, Maas,
  and Mendes}}]{Cucchieri:2006tf}
\bibinfo{author}{\bibfnamefont{A.}~\bibnamefont{Cucchieri}},
  \bibinfo{author}{\bibfnamefont{A.}~\bibnamefont{Maas}}, \bibnamefont{and}
  \bibinfo{author}{\bibfnamefont{T.}~\bibnamefont{Mendes}},
  \bibinfo{journal}{Phys.Rev.} \textbf{\bibinfo{volume}{D74}},
  \bibinfo{pages}{014503} (\bibinfo{year}{2006}).

\bibitem[{\citenamefont{Pascual and Tarrach}(1984)}]{Pascual:1984zb}
\bibinfo{author}{\bibfnamefont{P.}~\bibnamefont{Pascual}} \bibnamefont{and}
  \bibinfo{author}{\bibfnamefont{R.}~\bibnamefont{Tarrach}},
  \bibinfo{journal}{Lect. Notes Phys.} \textbf{\bibinfo{volume}{194}},
  \bibinfo{pages}{1} (\bibinfo{year}{1984}).

\bibitem[{\citenamefont{Davydychev et~al.}(1996)\citenamefont{Davydychev,
  Osland, and Tarasov}}]{Davydychev:1996pb}
\bibinfo{author}{\bibfnamefont{A.~I.} \bibnamefont{Davydychev}},
  \bibinfo{author}{\bibfnamefont{P.}~\bibnamefont{Osland}}, \bibnamefont{and}
  \bibinfo{author}{\bibfnamefont{O.~V.} \bibnamefont{Tarasov}},
  \bibinfo{journal}{Phys. Rev.} \textbf{\bibinfo{volume}{D54}},
  \bibinfo{pages}{4087} (\bibinfo{year}{1996}).

\bibitem[{\citenamefont{Ahmadiniaz and Schubert}(2013)}]{Ahmadiniaz:2012xp}
\bibinfo{author}{\bibfnamefont{N.}~\bibnamefont{Ahmadiniaz}} \bibnamefont{and}
  \bibinfo{author}{\bibfnamefont{C.}~\bibnamefont{Schubert}},
  \bibinfo{journal}{Nucl.Phys.} \textbf{\bibinfo{volume}{B869}},
  \bibinfo{pages}{417} (\bibinfo{year}{2013}).

\bibitem[{\citenamefont{Cornwall}(2012)}]{Cornwall:2012mk}
\bibinfo{author}{\bibfnamefont{J.~M.} \bibnamefont{Cornwall}}
  (\bibinfo{year}{2012}), \eprint{1211.2019}.

\bibitem[{\citenamefont{Jackiw and Johnson}(1973)}]{Jackiw:1973tr}
\bibinfo{author}{\bibfnamefont{R.}~\bibnamefont{Jackiw}} \bibnamefont{and}
  \bibinfo{author}{\bibfnamefont{K.}~\bibnamefont{Johnson}},
  \bibinfo{journal}{Phys. Rev.} \textbf{\bibinfo{volume}{D8}},
  \bibinfo{pages}{2386} (\bibinfo{year}{1973}).

\bibitem[{\citenamefont{Jackiw}(1973)}]{Jackiw:1973ha}
\bibinfo{author}{\bibfnamefont{R.}~\bibnamefont{Jackiw}}, \bibinfo{journal}{In
  *Erice 1973, Proceedings, Laws Of Hadronic Matter*, New York 1975, 225-251
  and M I T Cambridge - COO-3069-190 (73,REC.AUG 74) 23p}
  (\bibinfo{year}{1973}).

\bibitem[{\citenamefont{Cornwall and Norton}(1973)}]{Cornwall:1973ts}
\bibinfo{author}{\bibfnamefont{J.~M.} \bibnamefont{Cornwall}} \bibnamefont{and}
  \bibinfo{author}{\bibfnamefont{R.~E.} \bibnamefont{Norton}},
  \bibinfo{journal}{Phys. Rev.} \textbf{\bibinfo{volume}{D8}},
  \bibinfo{pages}{3338} (\bibinfo{year}{1973}).

\bibitem[{\citenamefont{Eichten and Feinberg}(1974)}]{Eichten:1974et}
\bibinfo{author}{\bibfnamefont{E.}~\bibnamefont{Eichten}} \bibnamefont{and}
  \bibinfo{author}{\bibfnamefont{F.}~\bibnamefont{Feinberg}},
  \bibinfo{journal}{Phys. Rev.} \textbf{\bibinfo{volume}{D10}},
  \bibinfo{pages}{3254} (\bibinfo{year}{1974}).

\bibitem[{\citenamefont{Poggio et~al.}(1975)\citenamefont{Poggio, Tomboulis,
  and Tye}}]{Poggio:1974qs}
\bibinfo{author}{\bibfnamefont{E.~C.} \bibnamefont{Poggio}},
  \bibinfo{author}{\bibfnamefont{E.}~\bibnamefont{Tomboulis}},
  \bibnamefont{and} \bibinfo{author}{\bibfnamefont{S.~H.~H.}
  \bibnamefont{Tye}}, \bibinfo{journal}{Phys. Rev.}
  \textbf{\bibinfo{volume}{D11}}, \bibinfo{pages}{2839} (\bibinfo{year}{1975}).

\bibitem[{\citenamefont{Schwinger}(1962{\natexlab{a}})}]{Schwinger:1962tn}
\bibinfo{author}{\bibfnamefont{J.~S.} \bibnamefont{Schwinger}},
  \bibinfo{journal}{Phys. Rev.} \textbf{\bibinfo{volume}{125}},
  \bibinfo{pages}{397} (\bibinfo{year}{1962}{\natexlab{a}}).

\bibitem[{\citenamefont{Schwinger}(1962{\natexlab{b}})}]{Schwinger:1962tp}
\bibinfo{author}{\bibfnamefont{J.~S.} \bibnamefont{Schwinger}},
  \bibinfo{journal}{Phys. Rev.} \textbf{\bibinfo{volume}{128}},
  \bibinfo{pages}{2425} (\bibinfo{year}{1962}{\natexlab{b}}).

\bibitem[{\citenamefont{Luscher and Weisz}(1995)}]{Luscher:1995vs}
\bibinfo{author}{\bibfnamefont{M.}~\bibnamefont{Luscher}} \bibnamefont{and}
  \bibinfo{author}{\bibfnamefont{P.}~\bibnamefont{Weisz}},
  \bibinfo{journal}{Nucl.Phys.} \textbf{\bibinfo{volume}{B452}},
  \bibinfo{pages}{213} (\bibinfo{year}{1995}).

\bibitem[{\citenamefont{Dashen and Gross}(1981)}]{Dashen:1980vm}
\bibinfo{author}{\bibfnamefont{R.~F.} \bibnamefont{Dashen}} \bibnamefont{and}
  \bibinfo{author}{\bibfnamefont{D.~J.} \bibnamefont{Gross}},
  \bibinfo{journal}{Phys. Rev.} \textbf{\bibinfo{volume}{D23}},
  \bibinfo{pages}{2340} (\bibinfo{year}{1981}).

\bibitem[{\citenamefont{Neuberger}(1987)}]{Neuberger:1986xz}
\bibinfo{author}{\bibfnamefont{H.}~\bibnamefont{Neuberger}},
  \bibinfo{journal}{Phys.Lett.} \textbf{\bibinfo{volume}{B183}},
  \bibinfo{pages}{337} (\bibinfo{year}{1987}).

\bibitem[{\citenamefont{Abbott}(1982)}]{Abbott:1981ke}
\bibinfo{author}{\bibfnamefont{L.~F.} \bibnamefont{Abbott}},
  \bibinfo{journal}{Acta Phys. Polon.} \textbf{\bibinfo{volume}{B13}},
  \bibinfo{pages}{33} (\bibinfo{year}{1982}).

\bibitem[{\citenamefont{Giusti et~al.}(2001)\citenamefont{Giusti, Paciello,
  Parrinello, Petrarca, and Taglienti}}]{Giusti:2001xf}
\bibinfo{author}{\bibfnamefont{L.}~\bibnamefont{Giusti}},
  \bibinfo{author}{\bibfnamefont{M.}~\bibnamefont{Paciello}},
  \bibinfo{author}{\bibfnamefont{C.}~\bibnamefont{Parrinello}},
  \bibinfo{author}{\bibfnamefont{S.}~\bibnamefont{Petrarca}}, \bibnamefont{and}
  \bibinfo{author}{\bibfnamefont{B.}~\bibnamefont{Taglienti}},
  \bibinfo{journal}{Int.J.Mod.Phys.} \textbf{\bibinfo{volume}{A16}},
  \bibinfo{pages}{3487} (\bibinfo{year}{2001}).

\bibitem[{\citenamefont{Taylor}(1971)}]{Taylor:1971ff}
\bibinfo{author}{\bibfnamefont{J.~C.} \bibnamefont{Taylor}},
  \bibinfo{journal}{Nucl. Phys.} \textbf{\bibinfo{volume}{B33}},
  \bibinfo{pages}{436} (\bibinfo{year}{1971}).

\end{thebibliography}

\end{document}